\documentclass{article}
\usepackage{amssymb}

\usepackage{graphicx}
\usepackage{amsmath}

\begin{document}

\title{On maximally supersymmetric Yang-Mills theories}
\author{M. Movshev\\Institut Mittag-Leffler \\Djursholm, Sweden\\ \\ A. 
Schwarz\thanks{The work of both authors was partially supported by NSF 
grant No. DMS 0204927}\\ Department of Mathematics\\ University of 
California \\ Davis, CA, USA} 
\date{\today}
\maketitle

\begin{abstract}
We consider 
ten-dimensional supersymmetric Yang-Mills theory (10D SUSY YM theory) 
and its dimensional reductions, in particular, BFSS and IKKT models.
We formulate these theories using algebraic techniques based on 
application of 
differential
graded Lie algebras and associative algebras as well as of 
more general 
objects, L$_{\infty}$- and A$_{\infty}$- algebras.

We show that using pure spinor formulation of 10D SUSY YM theory
equations of motion and isotwistor formalism one can interpret these 
equations as Maurer-Cartan equations for some differential Lie algebra. 
This statement can be used to write BV action functional of 10D SUSY YM
theory in Chern-Simons form. The differential Lie algebra we 
constructed is closely 
related to differential associative algebra $(\Omega, \bar {\partial})$
of $(0, k)$-forms on some supermanifold; 
the Lie algebra is  tensor product of $(\Omega, \bar {\partial})$ and matrix
algebra .
We construct several other algebras that are quasiisomorphic to 
$(\Omega, \bar {\partial})$ and, therefore, also can be used to give BV 
formulation of 10D SUSY YM theory and its reductions. In particular, 
$(\Omega, \bar {\partial})$ is quasiisomorphic to the algebra $(B, d)$, 
constructed by Berkovits. The algebras $({\Omega}_0, \bar {\partial})$ 
and 
$(B_0, d)$ obtained from $(\Omega, \bar {\partial})$ and $(B, d)$ by means 
of reduction to a point can be used to give a BV-formulation of IKKT 
model. 

We introduce associative algebra SYM as algebra where relations are 
defined as equations of motion of IKKT model and show that Koszul dual 
to the algebra $(B_0, d)$ is quasiisomorphic to SYM.
\end{abstract}

\section{ Introduction.}

The present paper begins a series of papers devoted to the analysis of the
ten-dimensional supersymmetric Yang-Mills theory (10D SUSY YM theory), its
dimensional reductions and its deformations having the same amount of
supersymmetries. Theories of this kind are quite important. Dimensional
reductions of 10D SUSY YM theory include 
the
reduction  to $\left(1+0\right) -$dimensional space first
considered in \cite{CH} and used in \cite{HN} to describe
supermembrane and in \cite{BFSS} (BFSS Matrix model ) to propose
mathematical formulation of M-theory
, IKKT Matrix 
model 
\cite{IKKT} (the reduction to a point) proposed as a nonperturbative
description of superstring , 
$N=4$ \ SUSY YM 
theory in four-dimensional
space. In some approximation dimensional reductions of 10D SUSY YM theory
describe fluctuations of $\ n$ \ coinciding flat D-branes; more precise
description of D-branes is provided by supersymmetric deformation of reduced
SUSY YM. We will analyze such deformations in \cite{MSch5}; this analysis can be used also to study Dirac-Born-Infeld (DBI) theory that can be considered as a SUSY deformation of SUSY YM where the terms containing derivations of field
strength are neglected.(In our discussion of  SUSY theories we always have in mind
maximally supersymmetric gauge theories.)

We will use algebraic techniques based on application of 
differential
graded Lie algebras and associative algebras as well as of 
more general 
objects, L$_{\infty}$- and A$_{\infty}$- algebras. 
In present paper we will give formulation of SUSY YM theories 
using these techniques. In the paper \cite{MSch3} the same 
techniques will be applied to more general gauge theories. 
The paper \cite{MSch4} will be devoted to calculation of some homology groups associated with the algebras considered in the present paper and in \cite{MSch3}. The consideration of supersymmetric deformations 
in \cite{MSch5} will be based on equivariant generalizations of results 
of \cite{MSch4}.

Main results of the present paper and of the subsequent 
papers \cite{MSch3}-\cite{MSch5} can be proven rigorously. 
However in these papers sometimes we give only heuristic 
arguments, avoiding pretty complicated mathematical proofs. 
Rigorous proofs will appear in \cite{M6}.

 Let us describe some of constructions we are using.
 
If $L$ is a differential graded Lie algebra we can write down the so called Maurer-Cartan equation 
\begin{equation}\label{E:mceq}
d\omega +\frac{1}{2}\left[ \omega , \omega \right] =0.
\end{equation}
This equation appears in numerous mathematical problems.
In particular, if \ $
L=\sum\limits_{k}\Omega ^{0, k}\otimes Mat_{n}=\Omega \otimes Mat_{n}$ \ is
the algebra of matrix-valued $\left(0, k\right) -$forms on complex manifold $
X, $ then the differential $\overline{\partial }=d\bar{x}^{k}\cdot \partial
/\partial \bar{x}^{k}$ specifies a structure of differential graded Lie
algebra on $L.$ (We can consider $\Omega $ \ and $L$ also as differential
graded associative algebras.) Solutions of the Maurer-Cartan equation (\ref{E:mceq})with $\omega \in \Omega ^{0, 1}\otimes Mat_{n}$ specify holomorphic
structures on topologically trivial vector bundle. We will use this remark
and isotwistor formalism \cite{Ros1}, \cite{GIKOS2}, \cite{SchwRos} to write equations of motion of 10D
SUSY YM as Maurer-Cartan equation for appropriate differential graded Lie
algebra. \ There are various ways to write these equations of motion in such
a form; such nonuniqueness  is related to the fact that quasiisomorphic
algebras have equivalent Maurer-Cartan equations. \ (A homomorphism of
differential algebras is called a quasiisomorphism if it generates an
isomorphism on homology.) \ The most convenient constructions are based on
the manifold \ ${\cal Q}=$ SO $\left(10, \mathbb{R}\right) /U\left(5\right) $ \
that can be interpreted as a connected component of the isotropic Grassmannian or a  space of pure
spinors. \ (The relation of pure spinors to 10D SUSY YM was discovered and
used in \cite{Howe1}, \cite{Howe2}, \cite{Berk3}). \ 

Maurer-Cartan equations are closely related to Chern-Simons action
functional; this general fact can be used to write down an action functional
of Chern-Simons type that describes \ 10D SUSY YM in BV formalism .

Maurer-Cartan equations and their generalizations to $L_{\infty }-$algebras
as well as corresponding Chern-Simons type action functionals play a very
important role in our constructions. \ It seems that Chern-Simons type
actions are ubiquitous; in some sense every action functional can be written
in Chern-Simons form (see Appendix C.)

We are basing our consideration on geometric version of Batalin-Vilkovisky
formalism \cite{SchwBV1}, \cite{SchwBV2}, \cite {AKSZ}. \ The main 
geometric notion we are using is the notion of \ 
$Q-$manifold (supermanifold equipped with an odd vector field obeying \ $
\left\{ Q, Q\right\} =0$). \ An algebraic counterpart of this notion is the
notion of \ $L_{\infty }-$algebra, that can be defined as a formal \ $Q-$
manifold. \ If a vector field \ $Q$ \ vanishes at \ $0, $ \ then its Taylor
expansion has the form 
\begin{equation}\notag
Q^{a}=Q_{b}^{a}x^{b}+Q_{bc}^{a}x^{b}x^{c}+\cdots
\end{equation}
The coefficients of this expansion can be used to define algebraic
operations \ $m_{1}, m_{2}, \ldots , $ \ that specify the structure of \ $
L_{\infty }-$algebra if \ $\left\{ Q, Q\right\} =0.$)

One can consider also non-commutative \ $Q-$manifolds
 (=$\mathbb{Z}_{2}$-graded 
associative algebras 
equipped with an 
odd derivation having square
equal to zero= differential $\mathbb{Z}_{2}$-graded 
associative algebras 
) and \ $A_{\infty }-$algebras 
(= formal non-commutative \ $Q-$
manifolds.) \ We prefer to work not with differential graded Lie algebras
and \ $L_{\infty }-$algebras but with differential graded associative
algebras and \ $A_{\infty }-$algebras. \ There is a standard way to obtain a
Lie algebra from associative algebras and \ $L_{\infty }-$algebra from \ $
A_{\infty }-$algebra; at the very end we apply this construction. \
Considering associative and \ $A_{\infty }-$algebras we are studying gauge
theories with all classical gauge groups at the same time.

We will give various formulations of 10D \ SUSY\ YM\ theory in terms of
differential associative algebras and \ $A_{\infty }-$algebras. \ 

One can use, for example, the algebra \ $\Omega =\Sigma \Omega ^{0, k}$ \ of
\ $\left(0, k\right) -$forms on complex (super) manifold \ 
$\mathcal R$ 
with
coordinates \ $\left(u^{\alpha }, \theta ^{\alpha }, z^{l}\right) $ \ where \ 
$u^{\alpha }$ \ is a pure spinor
 (i.e. \ $u^{\alpha }\Gamma _{\alpha \beta
}^{a}u^{\beta }=0$) \ 
and \ $$\left(u^{\alpha }, \theta ^{\alpha}, z^{l}\right)  \mbox{ is 
identified with }  \left( \lambda u^{\alpha},  
(\theta^{\alpha}-\varepsilon u^{\alpha} ), 
(z^{l}-\varepsilon \Gamma _{\alpha \beta}^{l}u^{\alpha}\theta^{\beta}) 
\right)$$ 
\ (Here \ $u^{\alpha }, \theta
^{\alpha }$ \ are ten-dimensional Weyl spinors, \ $z^{l}$ are coordinates in
ten-dimensional complex vector space, \ $\lambda $ \ is a non-zero complex
number, \ $\varepsilon $ \ is an odd parameter.)

The algebra \ $\Omega $ \ will be regarded as an associative algebra
equipped with differential \ $\bar{\partial}.$ \ The equations of motion of
10D SUSY YM theory in BV-formalism coincide with Maurer-Cartan equations for
differential Lie algebra corresponding to associative algebra \ $\Omega
\otimes Mat_{n}.$ \ The 
odd symplectic form of \ 
BV-formalism corresponds to
inner product on the algebra \ $\Omega .$ (\ Recall, that in \ BV-formalism
a classical system is described by a solution to the classical master
equation \ $\left\{ S, S\right\} =0$ \ on odd symplectic manifold. \
Geometrically it corresponds to an odd vector field \ obeying \ $\left\{
Q, Q\right\} =0$ \ and \ $L_{Q}\sigma =0$ \ where \ $\sigma $ \ is a closed
odd 2-form and \ $L_{Q}$ \ stands for Lie derivative. \ In the language of \ 
$L_{\infty }-$algebras the form \ $\sigma $ \ corresponds to an invariant
odd inner product on \ $L_{\infty }-$algebra. This inner product should be
non-degenerate on the homology of \ $L_{\infty }-$algebra.)

As we mentioned already we can use any differential graded algebra (or more
generally A$_{\infty }$-algebra) that is quasiisomorphic to $\Omega $ to
describe super Yang-Mills theory. In particular, we can use 
the algebra $B$
that consists of polynomial functions depending on pure spinor $u^{\alpha }, $
anticommuting spinor variable $\theta ^{\alpha }$ and 10 commuting variables 
$x^{0}, ..., x^{9}$. This algebra is equipped with a differential $d=u^{\alpha
}D_{\alpha }$ , where $D_{\alpha }=\frac{\partial }{\partial \theta ^{\alpha }
}-\Gamma _{\alpha \beta }^{m}\theta ^{\beta }\frac{\partial }{\partial x^{m}}
$. One can prove that differential algebra ($B, d$) is quasiisomorphic to
the differential algebra ($\Omega , \bar{\partial}$)(see 
Appendix A). The
action functional of 10D super Yang-Mills theory again can be written as a
Chern-Simons functional. The algebra $(B, d)$ and corresponding action
functional were introduced by Berkovits \cite{Berk3}. Notice that 
Berkovits gave  a formulation of superstring theory in terms
of pure spinors; the SUSY YM theory can be analyzed as massless sector
in the theory of open superstring \cite{Berk1},\cite{Berk2},
\cite{Gra1}, \cite{Gra2}.

The paper is organized as follows. We start with reminder of main facts
about ten-dimensional SUSY YM theory and about pure spinors (Sec. \ref{S:Maxx}). In
(Sec. \ref{S:isotw}) we use isotwistor formalism to write down equations of motion of \
10D SUSY YM theory in the form of Maurer-Cartan equations for algebra $
\Omega .$This allows us to present 10D SUSY YM theory by means of
Chern-Simons type action functional.

One can skip (Sec. \ref{S:isotw}) and go directly to (Sec. \ref{S:CH}) containing a description of
several quasiisomorphic differential algebras that can be used to analyze
10D SUSY YM theory and a construction of corresponding Chern-Simons type
action functionals. The constructions of (Sec. \ref{S:CH}) can be justified by means of
results of (Sec. \ref{S:isotw}), however, there exist other ways to do this. In
particular, one can justify these constructions using the language of \ $
A_{\infty }-$algebras; we sketch such a derivation in (Sec. \ref{S:gauge}).
Another way to justify our statements will be described in
\cite{MSch3}.
Appendices A and B contain some omitted proofs. The proofs of 
Appendix B are based on some general statements, that have also other 
interesting applications. Appendix C is devoted to the
exposition of some basic facts 
about 
L$_{\infty }$- and A$_{\infty }$-algebras, we follow mostly
\cite{AKSZ}, \cite{KS}, \cite{Keller} .
(This appendix, included for readers convenience, is
a part of paper \cite{MSch6}  containing review of the theory of
L$_{\infty }$- and A$_{\infty }$-algebras and some new
results.)

\section{  Maximally supersymmetric gauge theories.}\label{S:Maxx}

Let us start with consideration of ten-dimensional supersymmetric gauge
theory and its reductions. In components this theory is 
described by the
action functional

\begin{equation} \label{E:sym1}
S_{SYM}(A_{m}, \chi ^{a})=-\frac{1}{4}\mathrm{Tr}\, F_{mn}F^{mn}+\frac{1}{2}
\mathrm{Tr}\, \chi ^{\alpha }\Gamma _{\alpha \beta }^{m}[\mathcal{D}_{m}, \chi
^{\beta }]\,  
\end{equation}

where $A_{m}(x)$ and $\chi ^{\alpha }(x)$ are matrix valued 
functions on $\mathbb{R
}^{10}$ , $m=0, 1, \dots , 9$ is a vector index, $\alpha =1, \dots , 16$ is a
spinor index, $\mathcal{D}_{m}=\frac{\partial }{\partial x_{m}}+A_{m}(x)$
are covariant derivatives, $F_{nm}=\partial _{m}A_{n}-\partial
_{n}A_{m}+[A_{m}, A_{n}]$ stands for field strength. Notice, that in
ten-dimensional space left spinors $\chi ^{\alpha }$ and right spinors $
\varphi _{\alpha }$ are dual to each other ; the ten-dimensional $\Gamma $
-matrix takes the form 
\begin{equation}\notag
\Gamma _{m}=\left(
\begin{array}{cc}
0 & \Gamma _{m}^{\alpha \beta } \\ 
(\Gamma _{m})_{\alpha \beta } & 0
\end{array}
\right)
\end{equation}
The fermions $\chi ^{\alpha }$ are considered as anticommuting variables. We
consider our action functional $S$ as complex analytic functional of
fields. To quantize our theory we should integrate $e^{-S}$ over a real slice
in the complex space of fields. If we introduce Minkowski metric on $\mathbb{R
}^{10}$ we can single out the real slice imposing Hermiticity condition on $
A_{m}$ and Majorana condition on $\chi ^{\alpha }$ . With this choice the
real slice is invariant with respect to symmetries of the theory. It is
important to notice, however, that deforming the slice we don't change the
value of the functional integral. Therefore, there is no necessity to
respect the symmetries of the theory choosing the real slice. This remark is
important, for example, in the case of Euclidean metric on $\mathbb{R}^{10}$, 
when the notion of Majorana spinor is not defined. We will be interested
only in symmetries of the theory therefore we will always work with complex
analytic action functional; we have no necessity to discuss the choice of
real slice.

It will be convenient for us to work with holomorphic fields $A_{m}(x), {\chi 
}^{\alpha }(x)$ defined on complex space $V=\mathbb{C}^{10}$ and taking
values in $V^{\ast }\otimes Mat_{n}$ and $\Pi S\otimes Mat_{n}$
correspondingly. (Here $Mat_{n}$ stands for the algebra of complex matrices, 
$S$ denotes left spinor representation and $\Pi $ means parity reversion).

All our considerations will be local with respect to $x\in V.$ In other
words, our fields are polynomials or formal power series with respect to $x.$
This means, in particular, that action functionals we consider are formal
expressions (they contain an ill-defined integral over $x$)

The group of symmetries of action (\ref{E:sym1}) includes the two-sheeted
covering $Spin(10, \mathbb{C})$ of group $SO(10, \mathbb{C})$.

The action (\ref{E:sym1}) is also invariant under gauge transformations
parametrized by an endomorphism $\phi $: 
\begin{equation}\notag
V_{\phi }(A_{m})=[\mathcal{D}_{m}, \phi ]\, , \qquad V_{\phi }(\chi ^{\alpha
})=[\chi ^{\alpha }, \phi ]\, .
\end{equation}

The action functional (\ref{E:sym1}) is invariant under the supersymmetry
transformations 
\begin{eqnarray}
\delta _{\epsilon }(A_{m}) &=&\epsilon ^{\alpha }(\Gamma _{m})_{\alpha \beta
}\chi ^{\beta }\, ,  \notag  \\
\delta _{\epsilon }(\chi ^{\alpha }) &=&\frac{1}{2}{(\Gamma ^{mn})^{\alpha }}
_{\beta }\epsilon ^{\beta }F_{mn}\notag
\end{eqnarray}
as well as under trivial supersymmetry transformations 
\begin{equation}\notag
\tilde{\delta}_{\epsilon }(A_{m})=0\, , \quad \tilde{\delta}_{\epsilon }(\chi
^{\alpha })=\epsilon ^{\alpha }\, .
\end{equation}
Each of the last two transformations is parametrized by $\epsilon ^{\alpha }$ -
a constant Weyl spinor, i.e. a spinor proportional to the unit endomorphism.

The equations of motion corresponding to (\ref{E:sym1}) (but not the action
functional itself) can be written in explicitly supersymmetric form in $
(10|16)$ -dimensional superspace $\mathbb{C}^{10|16}$. To do this we define
a connection on this space as a collection of covariant derivatives $
D_{m}=\partial _{m}+A_{m}(x, \theta)$ , $D_{\alpha }=D_{\alpha }+A_{\alpha
}(x, \theta)$ where $\partial _{m}=\frac{\partial }{\partial x^{m}}$, $
D_{\alpha }=\frac{\partial }{\partial {\theta }^{\alpha }}-{{\Gamma }
_{\alpha \beta }}^{m}\theta ^{\beta }\partial _{m}$ , $A_{m}$ and $A_{\alpha
} $ are holomorphic matrix functions of ten-dimensional vector $x^{m}$ and
Weyl spinor $\theta ^{\alpha }$ (we consider $\theta ^{\alpha }$ as
anticommuting variables). The curvature of the connection $D=(D_{m}, D_{\alpha
})$ has components 
\begin{eqnarray}
&&F_{\alpha \beta }=\{\mathcal{D}_{\alpha }, \mathcal{D}_{\beta }\}+2\Gamma
_{\alpha \beta }^{m}\mathcal{D}_{m}\, ,  \notag \\
&&F_{\alpha m}=[\mathcal{D}_{\alpha }, \mathcal{D}_{m}]\, , \qquad F_{mn}=[
\mathcal{D}_{m}, \mathcal{D}_{n}]\, .
\end{eqnarray}

We will say that a connection $D$ is a gauge superfield if it satisfies the
constraint 
\begin{equation}\label{curv}
F_{\alpha \beta }=0
\end{equation}
It is easy to check that the set of gauge superfields is invariant with
respect to supergauge transformations 
\begin{equation}\label{supergauge}
A_{m}\mapsto g^{-1}A_{m}g+g^{-1}\mathcal{D}_{m}g\, , \quad A_{\alpha }\mapsto
g^{-1}A_{\alpha }g+g^{-1}\mathcal{D}_{\alpha }g\, . 
\end{equation}
and supersymmetry transformations 
\begin{equation}\label{ss}
\delta _{\gamma }(\mathcal{D}_{m})=[\mathcal{D}_{m}, \tilde{D}_{\gamma
}]\, , \quad \delta _{\gamma }(\mathcal{D}_{\alpha })=\{\mathcal{D}_{\alpha }, 
\tilde{D}_{\gamma }\}\, .
\end{equation}
where
\begin{equation} \label{tildeD}
\tilde D_{\alpha} = \frac{\partial}{\partial \theta^{\alpha}} + \Gamma^{m}_{\alpha \beta}\theta^{\beta} \partial_{m} \, .
\end{equation}

One can identify the 
moduli space of gauge superfields (the set of 
solutions
of (\ref{curv}) factorized with respect to supergauge transformations) with
the moduli space of solutions to the equations 
\begin{equation}\label{eqsofm}
\begin{split}
&\Gamma _{\alpha \beta }^{m}\mathcal{D}_{m}\chi ^{\beta }=0  \\
&\mathcal{D}_{m} F^{mn}=\frac{1}{2}\Gamma _{\alpha \beta}^{n}\{\chi ^{\alpha }, \chi ^{\beta }\}
\end{split}
\end{equation}
(space of solutions of the equations of motion corresponding to the action
functional (\ref{E:sym1}) up to gauge transformations.) To prove this statement
one can notice that the gauge condition 
\begin{equation}\label{gauge}
\theta ^{\alpha }A_{\alpha }=0 
\end{equation}
restricts the group of supergauge transformation to the group of gauge
transformations. All components of gauge superfield $(\mathcal{A}_{m}, 
\mathcal{A}_{\alpha })$ obeying gauge condition (\ref{gauge}) can be
restored by means of some recurrent process \cite{Zhel} from the zero components of
superfields $A_{m}$ and $\chi ^{\alpha }={(}${$\Gamma $}${^{m})}^{\alpha
\beta }F_{\beta m}$. These components are identified with corresponding
fields in the component formalism; they obey the equation (\ref{eqsofm}).

Notice that using (\ref{curv}) one can express $A_{m}(x, \theta)$ in terms
of $A_{\alpha }(x, \theta)$. This remark permits us to write the equation of
motion in superfield formalism as a condition
\begin{equation}\notag
\{D(u), D(u)\}=0, 
\end{equation}
where $D(u)=u^{\alpha }D_{\alpha }=u^{\alpha }(D_{\alpha }+A_{\alpha })$ and 
$u^{\alpha }$ is an arbitrary none-zero spinor obeying $u^{\alpha }{{\Gamma }
_{\alpha \beta }}^{m}u^{\beta }=0$ (in other words $u^{\alpha }$ are
ten-dimensional pure spinor). This pure spinor formulation of equations of motion \cite{Howe1}, \cite{Howe2} can be used to apply the ideas of isotwistor formalism \cite{Howe1}, \cite{GIKOS2}, \cite{SchwRos}.
The manifold of pure spinors $\tilde{{\cal Q}}$ and corresponding projective
manifold ${\cal Q}$ play the main role in this formulation. (In ${\cal Q}$ we identify
spinors $u^{\alpha }$ and $\lambda u^{\alpha }$. In other words ${\cal Q}$ is
obtained from $\tilde{{\cal Q}}$ by means of factorization with respect to vector
field $E=u^{\alpha }\frac{\partial }{\partial u^{\alpha }}$). It is clear
that ${\cal Q}$ and $\tilde{{\cal Q}}$ are complex manifolds, 
${dim}_{\mathbb{C}}{\cal Q}=10$, ${
dim}_{\mathbb{C}}\tilde{{\cal Q}}=11$. The group $SO(10, \mathbb{R})$ acts as on ${\cal Q}$
transitively with stable subgroup $U(5)$. This means that we can identify ${\cal Q}$
with $SO(10, \mathbb{R})/U(5)$. The manifold $\tilde{{\cal Q}}$ can be considered as
a total space of $\mathbb{C}^{\ast }$ bundle $\alpha $ over ${\cal Q}$ or as a
homogeneous space $Spin(10, \mathbb{R})\times \mathbb{C}^{\ast }/\tilde{U}(5)$
.(Here ${\ \mathbb{C}}^{\ast }$ stands for the multiplicative group of
non-zero complex numbers and $\tilde{U}(5)$ is a two-sheeted cover of $U(5)$
).

Sometimes it is convenient to use instead of $\widetilde{{\cal Q}}$ the manifold $
\widehat{{\cal Q}}=Spin(10, \mathbb{R})\times \mathbb{C}/\widetilde{U}(5)$ (the
total space of $\mathbb{C}$-bundle $\widehat{\alpha }$ over ${\cal Q}$).

The complex group $Spin(10, \mathbb{C})$ also acts transitively on ${\cal Q}$ ; 
corresponding stable subgroup $P$ is a parabolic subgroup. To describe the
Lie algebra $\mathfrak{p}$ of $P $ we notice that the Lie algebra $
\mathfrak{so}(10, \mathbb{C}) $ of $SO(10, \mathbb{C}) $ can be identified
with ${\Lambda}^2(V) $ (with the space of antisymmetric tensors ${\rho}_{a
b} $ where $a, b=0, \dots, 9 $). The vector representation $V$ of $SO(10, 
\mathbb{C}) $ restricted to the group $GL(5, \mathbb{C}) \subset SO(10, 
\mathbb{C}) $ is equivalent to the direct sum $W \oplus W^* $ of vector and
covector representations of $GL(5, \mathbb{C}) $. The Lie algebra of $SO(10, 
\mathbb{C}) $ as vector space can be decomposed as ${\Lambda}^2(W)\oplus 
\mathfrak{p} $ where $\mathfrak{p }=(W \otimes W^*) \oplus \Lambda^2(W^*) $
is the Lie subalgebra of $\mathfrak{p}$. Using the language of generators we
can say that the Lie algebra $\mathfrak{so}(10, \mathbb{C}) $ is generated by
skew-symmetric tensors $m_{ab}, n^{ab} $ and by $k_a^b $ where $
a, b=1, \dots, 5 $. The subalgebra $\mathfrak{p}$ is generated by $k_a^b $
and $n^{ab} $. Corresponding commutation relations are 
\begin{align}
& [m, m^{\prime}]=[n, n^{\prime}]=0 \\
&[m, n]_a^b=m_{ac}n^{cb} \\
&[m, k]_{ab}=m_{ac}k_b^c+m_{cb}k_a^c \\
&[n, k]_{ab}=n^{ac}k_c^b+n^{cb}k_c^a
\end{align}

\section{ Isotwistor formalism. }\label{S:isotw}

To apply the ideas of isotwistor formalism to equations of motion
in the form (\ref{curv}) we need some information about holomorphic vector bundles
on the manifold ${\cal Q}$ of pure spinors. We will assume that every topologically
trivial holomorphic vector bundle over ${\cal Q}$ is holomorphically trivial. This
statement can be proved for semistable bundles \footnote{ A. Tyurin, private
communication}; holomorphic triviality of small deformation of
holomorphically trivial bundle immediately follows from $H^{0, 1}({\cal Q})=0.$

Let as introduce a supermanifold $\tilde{{\cal R}}$ as a direct product of $\tilde{{\cal Q}
}$, $V$ and $\Pi S$. Here $V$ stands for the space of vector representation
of $SO(10, \mathbb{C})$ and $S$ for the space of its spinor representation.
(We are working always with left spinors.) The manifold $\tilde{{\cal R}}$ can be
interpreted also as a quotient $G\times \mathbb{C}^{\ast }/P$ where $G$
stands for complex super Poincare group and $P\subset Spin(10, \mathbb{C})$
is embedded in natural way into $G\times {\mathbb{C}}^{\ast }$. A point of $
\tilde{{\cal R}}$ can be represented as a triple $(u^{\alpha }, x^{m}, {\theta }
^{\alpha })$ where $u^{\alpha }$is a pure spinor , $x^{m}\in V$, $\theta
^{\alpha }\in \Pi S$. We define an odd vector field on $\tilde{{\cal R}}$ by the
formula 
\begin{equation}\notag
D=u^{\alpha }D_{\alpha }=u^{\alpha }(\frac{\partial }{\partial \theta
^{\alpha }}+{\Gamma _{\alpha \beta }}^{m}\theta ^{\beta }\frac{\partial }{
\partial x^{m}})
\end{equation}
It satisfies $D^{2}=0$. A complex manifold ${\cal R}$ will be defined as a quotient
of $\tilde{{\cal R}}$ with respect to vector fields $D$ and $$E=u^{\alpha }\frac{
\partial }{\partial u^{\alpha }}$$ The vector fields $D$ and $E$ commute
with the action of super Poincare group $G$. This means that ${\cal R}$ can be
considered as a quotient of $G$ with respect to a subgroup generated by a
semidirect product of $P$ and a subgroup of dimension $(0|1)$ .(If $S$ is
represented as a fermionic Fock space the generator of this subgroup
corresponds to the vacuum.) It will be convenient to consider also the
manifold ${\cal R}^{\prime }$ obtained from $\tilde{{\cal R}}$ by means of factorization
with respect to vector field $E$. This manifold is intermediate between $
\tilde{{\cal R}}$ and ${\cal R}$; to obtain ${\cal R}$ from ${\cal R}^{\prime }$ we factorize with
respect to $D$.

The manifold ${\cal R}$ is homogeneous with respect to the action of super-Poincare
group.

Let us consider now the algebra $\Omega$ of $(0, k)$-forms on the
(super) manifold ${\cal R}$. This is a supercommutative associative algebra equipped
with differential $\bar{\partial}$. Introduce  also 
\begin{equation}\label{alg}
\Omega \otimes Mat_{n} 
\end{equation}
(the algebra of matrix-valued $(0, k)$-forms). Considering (\ref{alg}) as a
differential Lie algebra we can write down the corresponding Maurer-Cartan
equation $\bar{\partial}\omega +\frac{1}{2}\{\omega , \omega \}=0$. We will
show that in the case when $\omega $ is a $(0, 1)$-form this equation is
equivalent to (\ref{curv}). First of all, we notice that $\Omega \otimes Mat_{n}$ is a subalgebra of the algebra of $(0, k)$
-forms on $\tilde{{\cal R}}$. More precisely, vector fields $D, E$ and their complex
conjugate fields $\bar{D}, \bar{E}$ determine operators $L_{D}, L_{\bar{D}
}, i_{D}, i_{\bar{D}}, L_{E}, L_{\bar{E}}, i_{E}, i_{\bar{E}}$ (Lie derivatives
and contractions). A form $\omega $ on $\tilde{{\cal R}}$ descends to ${\cal R}$ if 
\begin{equation}\label{E:asjhgv}
\begin{split}
& L_{D}\omega =L_{\bar{D}}\omega =i_{D}\omega =i_{\bar{D}}\omega =0 \\
& L_{E}\omega =L_{\bar{E}}\omega =i_{E}\omega =i_{\bar{E}}\omega =0
\end{split}
\end{equation}
(We work with $(0, k)$-forms hence $i_{D}\omega , i_{E}\omega $ always vanish.)

We will consider a matrix valued $(0, 1)$-form $\omega $ on $\tilde{{\cal R}}$ that
obeys (\ref{E:asjhgv}) and satisfies Maurer-Cartan equations. It descends to ${\cal R}$ and
to ${\cal R}^{\prime }$. On ${\cal R}^{\prime }$ it defines a topologically trivial
holomorphic vector bundle; such a bundle is also holomorphically trivial.
(One can derive this statement from corresponding statement for manifold ${\cal Q}.$
) This means that there exists a matrix-valued function $g$ on $\tilde{{\cal R}}$
that obeys $\omega =g^{-1}\bar{\partial}g$ and descends to ${\cal R}^{\prime }$
(satisfies $Eg=\bar{E}g=0).$ The operator $gDg^{-1}$ can be written in the
form 
\begin{equation}\notag
gDg^{-1}=D+u^{\alpha }A_{\alpha }=u^{\alpha }\left(D_{\alpha }+A_{\alpha
}\right) .
\end{equation}
It follows from $\left\{ gDg^{-1}, gDg^{-1}\right\} =0$ that $A_{\alpha }$
obeys the Yang-Mills equation of motion. From the other side it follows from
(\ref{curv}) that for every solution of YM equations of motion $A_{\alpha }\left(
x, \theta \right) $ there exists such a matrix-valued function $g$ on \ $
\tilde{{\cal R}}$ \ that 
\begin{equation}\notag
gDg^{-1}=D\left(u\right) =u^{\alpha }\left(D_{\alpha }+A_{\alpha }\right)
\end{equation}
moreover,  function $g$ can be chosen in such a way that 
\begin{equation}\notag
g^{-1}\cdot \overline{D}\cdot g=\overline{D}, \quad g^{-1}\cdot E\cdot
g=E, \quad g^{-1}\cdot \overline{E}\cdot g=\overline{E}, 
\end{equation}
or, in other words, 
\begin{equation}\notag
\overline{D}g=0, \quad Eg=0, \quad \overline{E}g=0.
\end{equation}
Then the form \ $\omega =g^{-1}\overline{\partial }g$ obeys Maurer-Cartan
equations and the conditions (\ref{E:asjhgv}).

\section{Various formulations of 10D SUSY YM.\\ Chern-Simons action functionals.}\label{S:CH}

We have shown how one can express the SUSY YM equations of motion in the
form of Maurer-Cartan equations of motion for the algebra \ $\Omega \otimes
Mat_{n}.$ Recall that $\Omega =\sum \Omega ^{0, k}$\ stands for the algebra
of $(0, k)$-forms on complex supermanifold ${\cal R}.$ (We will use the notation $
\Omega (M)$ for the algebra of $(0, k)$-forms on a manifold $M; $ hence $\Omega =\Omega
({\cal R})$.) As we mentioned in the introduction many important theories can be
obtained by means of dimensional reduction of 10D SUSY YM; 
in 
particular, \ IKKT Matrix \ model is obtained by means of
reduction to a 
point.\ Of
course, all of these theories can be obtained also by means of dimensional
reduction of action functional (\ref{E:sym1}). For example, IKKT Matrix model
can be described by means of the algebra \ $\Omega _{0}$ \ (dimensional
reduction of \ $\Omega)$ . More precisely, the algebra \ $\Omega _{0}$ is
defined as $(\Omega ({\cal R}_{0}), \overline{\partial })$ where ${\cal R}_{0}$ is obtained
from $\widetilde{{\cal R}}_{0}=\widetilde{{\cal Q}}\times \Pi S$ by means of factorization
with respect to vector fields $E=u^{\alpha }\frac{\partial }{\partial
u^{\alpha }}+ \theta^{\alpha }\frac{\partial }{\partial
 \theta^{\alpha }}$ and $d=u^{\alpha }\frac{\partial }{\partial \theta ^{\alpha }}$.

We will discuss now BV-formalism of reduced theory in terms of algebra $
\Omega _{0}; $ the generalization to complete 10D SUSY YM theory is
straightforward. \bigskip

It will be essential later that ${\cal R}_{0}$ is a split supermanifold (a
supermanifold that can be obtained from a vector bundle by means of parity
reversion on fibers). Moreover, the manifold ${\cal Q}$ 
(underlying manifold for ${\cal R}_{0}$
) is a complex manifold equipped with a transitive holomorphic action of the
group $SO(10, \mathbb{R})$ with a stabilizer $U(5)$. ${\cal R}_{0}$ can be
obtained from a homogeneous vector bundle corresponding \footnote{
Recall that for a representation of $H$ on vector space $V$ we can construct
a homogeneous vector bundle $G/H$ having $(G\times V)/H$ as the total space
and $V$ as a fiber.} to two-valued representation $T$ of $U(5)$. Here $T 
$ is $(\Lambda ^{2}(W)+\Lambda ^{4}(W))\otimes det^{-\frac{1}{2}}(W)$ where $
W$ stands for vector representation of $U(5).$

Let us write down the BV action functional for reduced 10D SUSY\ YM theory
in terms of $\Omega _{0}$ as a generalized Chern-Simons action functional.

Namely, the action functional is defined on the algebra \ $\Omega
_{0}\otimes Mat_{n}$ by \ the formula

\begin{equation}\label{E:chsm}
f\left(\omega \right) =Tr\left(\frac{1}{2}\omega \bar{\partial}
\omega +\frac{2}{3}\omega \left[ \omega, \omega \right] \right)
\end{equation}
\bigskip

where the functional \ $Tr$ \ does not vanish only on the forms of degree 3.

The construction of functional $Tr$ involves integration. In the integration
theory on supermanifolds one should use so called integral forms \cite{BL} instead of
differential forms. An integral $k$ -form can be defined as an expression 
\begin{equation}\notag
\psi =C^{A_{1}\dots A_{k}}(z)\rho (z)
\end{equation}
where $z^{A}$ are (even and odd) coordinates on a supermanifold, $\rho (z)$
stands for volume element, $C^{A_{1}\dots A_{k}}$ is a (super) antisymmetric
tensor of rank $k$. In mathematical terms $\rho $ is a section of Berezinian
line bundle $Ber$ (a bundle having (super)Jacobians as transition
functions) and coefficients $C^{A_{1}\dots A_{k}}$ are sections of $\Lambda
^{k}(T)$ where $T$ stands for tangent bundle and $\Lambda ^{k}$ is an
exterior power in the sense of superalgebra \footnote{Recall that for $\mathbb{Z}_{2}$-graded space $T=T_{0}+T_{1}$ the (super)exterior algebra $\Lambda ^{\bullet }(T)$ is $\Lambda ^{\bullet}(T_{0})\otimes S^{\bullet }(T_{1})$ where $S^{\bullet }$ stands for symmetric algebra}.

Suppose we are given a super-manifold of dimension $(a, b), $ an integral form 
$\psi $ of degree $k$ and immersed super-manifold $X$ of dimension $(a-k, b)$
. Then there is a well defined operation of integration of the form $\psi $
over submanifold $X$. In other words immersed submanifolds of right
dimension define functionals on integral forms. There is a different method
to define functionals on the space $\Omega _{int}^{-k}$ of  integral $k$
-forms. Take a differential form $\lambda $ of degree $k$. After making
contractions of vector and covector indexes one gets a section $<\psi
, \lambda >$ of $Ber$ which can be integrated over the manifold. Hence we
obtain a pairing between integral and differential forms. One can introduce
an analog of de Rham differential on the space of integral forms and to
prove an analog on Stokes' formula. The differential is compatible with the
pairing of differential and integral forms.

On a complex manifold a real line bundle $Ber$ is a tensor product $Ber_{\mathbb C}\otimes \overline{Ber}_{\mathbb C}$, where $Ber_{\mathbb C}$ is a holomorphic line bundle, called a holomorphic Berezinian, the bar symbol means a complex conjugation. The symbol $\otimes$ in context of vector bundles will always mean their fiberwise  tensor product.

A complexification of the tangent vector bundle $T$ 
of a complex manifold splits into a sum 
$T_{\mathbb C}+\overline{T}_{\mathbb C}$, where  $T_{\mathbb C}$ 
is a holomorphic and $\overline{T}_{\mathbb C}$ is an antiholomorphic 
component. This implies that
$$\Omega _{int}^{-k}=Ber\otimes 
\Lambda^{k}(T)=\bigoplus_{i+j=k} Ber_{\mathbb C}\otimes \Lambda^{i}(T_{\mathbb C})\otimes \overline{Ber}_{\mathbb C} \otimes \Lambda^{j}(\overline{T}_{\mathbb C})$$
In other words, on complex manifold we can consider not only
differential $(i,j)$-forms, but also integral  $(i,j)$-forms.

The integral form we about to construct is a $(0,3)$-form, i.e. it lives 
in $Ber_{\mathbb 
C}\otimes \overline{Ber}_{\mathbb C} \otimes 
\Lambda^{3}(\overline{T}_{\mathbb C}) .$
In the case at hand we are working with a split supermanifold ${\cal R}_0$. The space $(\Lambda^3(W)\oplus \Lambda^1(W))\otimes det^{-1}(W)\otimes \overline{\Lambda^3(W)\otimes det^{-1}(W)}$ contains a nontrivial $U(5)$
invariant element, defined up to a constant. If one takes a seventh exterior
power of it, one gets an element 
\begin{equation} \label{E:hdohn}
tr \in \Lambda^7(\Lambda^3(W)\oplus \Lambda^1(W))\otimes det^{-7}(W)\otimes 
\overline{\Lambda^7(\Lambda^3(W))\otimes det^{-7}(W)}
\end{equation}
We will use this element to construct an $SO(10)$-invariant integral 
form.
Smooth sections of holomorphic Berezinian $Ber_{\mathbb C}$ can be 
identified with
sections of a homogeneous vector bundle over ${\cal Q}$ corresponding  
to the following representation of $U(5)$ 
\begin{equation}\label{E:fsdg}
{\cal B}=A^{\bullet}\otimes \overline{A}^{\bullet}\otimes 
det^{-\frac{7}{2}}(W)
\end{equation}
where $$A^{\bullet}=\Lambda^{\bullet}[(\Lambda^3(W)\oplus \Lambda^1(W))\otimes det^{-\frac{1}{2}}(W)]$$

The smooth sections of vector bundle over ${\cal Q}$ that 
corresponds to $A^{\bullet}\otimes \overline{A}^{\bullet}$ can be 
identified with the ring of smooth superfunctions on the manifold ${\cal 
R}_0$.

 Sections of vector bundles over ${\cal R}_0$ are modules over the ring 
of superfunctions on ${\cal R}_0$. Sections of homogeneous vector bundles 
over ${\cal R}_0$ can be interpreted as sections of some homogeneous 
vector bundles over ${\cal Q}$, which in turn are induced from $U(5)$ representations. These  representations  are modules over $A^{\bullet}\otimes \overline{A}^{\bullet}$-the ring which is the substitute  in the representation theory for  the ring of 
superfunctions on ${\cal R}_0$. In the following if representations $X$ 
and $Y$ are modules over $A^{\bullet}\otimes \overline{A}^{\bullet}$, we denote their tensor product over $A^{\bullet}\otimes \overline{A}^{\bullet}$ as $X\underset{A^{\bullet}\otimes \overline{A}^{\bullet}}{\otimes} Y$. Such tensor product of representations corresponds to the tensor product of vector bundles.

The line bundle $Ber_{\mathbb C}$ over ${\cal R}_0$ is a pullback of a 
homogeneous line 
bundle $L$ over  ${\cal Q}$, that  corresponds to $U(5)$ representation 
$det^{-\frac{7}{2}}(W)$ (this is a local computation). 



The third exterior power of tangent bundle to ${\cal R}_0$ corresponds to
representation 
\begin{equation}\notag
\Lambda^3(T)=\left(\bigoplus_{i=0}^{3}\Lambda^i(\Lambda^2(W))\otimes S^{3-i}(\Lambda^2(W)\oplus \Lambda^4(W))\otimes det^{-\frac{3-i}{2}}(W)\right) \otimes A^{\bullet}\otimes \overline{A}^{\bullet}
\end{equation}

Smooth  integral $(0,3)$-forms  on manifold ${\cal 
R}_0$ can be considered as sections 
of a vector bundle induced from representation $$ {\cal 
B}\underset{A^{\bullet}\otimes \overline{A}^{\bullet}}{\otimes} 
\overline{{\cal 
B}\underset{A^{\bullet}\otimes \overline{A}^{\bullet}}{\otimes} \Lambda^{3}
(T_{{\cal R}})}.$$ It is clear that $\overline{\Lambda ^{7}(\Lambda 
^{3}(W))\otimes det^{-7}(W)}$ can be identified with 
\begin{align}
&\overline{\Lambda ^{3}(\Lambda ^{2}(W))\otimes det^{-\frac{8}{2}}(W)}\subset \notag \\
&\subset \overline{\Lambda ^{3}(\Lambda ^{2}(W))\otimes \Lambda ^{15}[(\Lambda ^{3}(W)\oplus \Lambda ^{1}(W))\otimes det^{-\frac{1}{2}}(W)]\otimes det^{-\frac{7}{2}}(W)}\subset \notag \\
&\subset \overline{ {\cal B}\underset{A^{\bullet}\otimes 
\overline{A}^{\bullet}}{\otimes} \Lambda ^{3}(T_{{\cal R}})}.\notag
\end{align}
We see that $tr$ introduced in (\ref{E:hdohn}) defines a 
$U(5)$-invariant 
element 
of $ {\cal B}\underset{A^{\bullet}\otimes 
\overline{A}^{\bullet}}{\otimes} \overline{ {\cal 
B}\underset{A^{\bullet}\otimes \overline{A}^{\bullet}}{\otimes}\Lambda ^{3}(T_{{\cal R}})}$  . The 
corresponding $SO(10)$- invariant section is the integral form we are 
looking for. We denote this section by the same symbol $tr$.
It is easy to prove that the element $tr\in Ber_{\mathbb C}\otimes \Lambda^{3}(\overline{T}_{\mathbb C}) \otimes \overline{Ber}_{\mathbb C} \subset \Omega _{int}^{-3}$ is a $d$ closed integral form. It can be used
to define a functional of the space of differential $(0, 3)$-forms on
manifold ${\cal R}_{0}$ by the formula 
\begin{equation}\notag
Tr(a)=\int_{{\cal R}_{0}}<tr, a>
\end{equation}
Using the functional $Tr$ we define an odd 2-form on $\Omega _{0}:$
\begin{equation}\notag
\sigma (a, b)=Tr(a\wedge b)
\end{equation}
This 2-form defines an odd $\overline{\partial }$-invariant presymplectic
structure on $\Omega _{0}\otimes Mat_{n}.$ The functional (\ref{E:chsm}) obeys master
equation with respect to this presymplectic structure. (More precisely, it
descends to corresponding odd symplectic manifold and satisfies master
equation there.) This follows from the fact that the equations of motion
corresponding to (\ref{E:chsm}) are Maurer-Cartan equations 
 coming from
an odd vector field having square equal to zero.

To write down an odd presymplectic structure on $\Omega _{0}\otimes Mat_{n}$
and BV action functional for 10D SUSY YM theory we use the same formulas as
in reduced case adding everywhere integration with respect to spatial
coordinates.

Notice, that the formulation of SUSY YM and its reductions in terms of
the algebra of $(0,k)$-forms is  similar to the formulation
of B-model on Calabi-Yau manifold in terms of holomorphic Chern-Simons
action functional. The algebra of B-model is the same, but the odd
symplectic structure is different. One can say that we formulated 
SUSY YM as a generalized B-model on a supermanifold.

Let us describe several algebras that are quasiisomorphic to $\Omega $ and $
\Omega _{0}$ and, therefore can be used to analyze 10D SUSY YM and its
reduction to a point.

\bigskip The algebra $B_{0}$ is defined as an algebra of polynomial
functions of pure spinor $u^{\alpha }$ and anticommuting spinor $\theta^{\alpha }$. It is equipped with differential 
\begin{equation}\notag
d=u^{\alpha }\frac{\partial }{\partial \theta ^{\alpha }}.
\end{equation}

One can say that $B_{0}$ is a tensor product of the algebra $F(\widetilde{{\cal Q}}
) $ of polynomial functions on the manifold $\widetilde{{\cal Q}}$ and exterior
algebra $\Lambda (S^{\ast })=F(\Pi S).$ The algebra $B_{0}$ can be obtained
by means of reduction to a point from the algebra $B=B_{0}\otimes F(V)$
equipped with differential 
\begin{equation}\notag
d=u^{\alpha }(\frac{\partial }{\partial \theta ^{\alpha }}+\Gamma _{\alpha
\beta }^{m}\theta ^{\beta}\frac{\partial }{\partial x^{m}})
\end{equation}
where $F(V)$ stands for the space of functions depending of $
x^{1}, ..., x^{m}. $

One can prove that the algebra $B$ is quasiisomorphic to $\Omega $ and $
B_{0} $ is quasiisomorphic to $\Omega _{0}$ (see Appendix A.) This statement
permits us to formulate 10D SUSY YM theory in terms of algebra $B$ and its
reduction to a point (IKKT model) in terms of $B_{0}.$ The action functional
again has Chern-Simons form 
\begin{equation}\label{E:chsm2}
f\left(\omega \right) =Tr\left(\frac{1}{2}\omega d
\omega +\frac{2}{3}\omega \left[ \omega, \omega \right] \right)
\end{equation}
where $Tr$ stands for the functional on $B_{0}$ that is determined by the
following properties. Let us define the space $B^{5, 3}_{0}$ 
as a subspace 
of $B_{0}$, generated by monomials 
$u^{\alpha_1 }\dots u^{\alpha_3 } \theta ^{\beta_1 }\dots 
\theta ^{\beta_5 }$. It contains a unique (up 
to a factor) $Spin(10)$- invariant
non-zero vector 
$\lambda$. Denote a $Spin(10)$- invariant projection on 
$\lambda$ by $p$. Then by definition 
$$Tr(\omega)\lambda=p(\omega).$$
The functional $Tr$ is 
unique up to multiplication on non-zero constant.  (We 
considered here the action functional of IKKT
model; to obtain the action functional of full 10D SUSY YM theory one should
work with algebra $B$ and include integration over $\mathbb{R}^{10}$ in the
definition of action functional.) \bigskip

One more description of IKKT Matrix model can be obtained from isotwistor
formulation \cite{SchwRos} of \ $N=3$ \ four-dimensional \ SUSY\ YM theory (that is
equivalent to its \ $N=4$ \ counterpart after reduction to a point). \ This description is
based on the following algebra \ $R_{N=3}$ described below. The algebra $R_{N=3}$  is quasiisomorphic to \ $
\Omega _{0}$ \ and \ $B_{0}.$

Let $k[p_{1}, \dots , p_{3}, u^{1}, \dots , u^{3}, z_{1}, z_{2}, w_{1}, w_{2}]$ be a
polynomial algebra. The Lie group $GL(3)\times SU_{L}(2)\times SU_{R}(2)$
acts on linear space $W$ spanned by generators $p_{1}, \dots , p_{3}, u^{1}, \dots , u^{3}, z_{1}, z_{2}, w_{1}, w_{2}$. The representation $W$
splits into a sum of irreducibles 
\begin{equation}\notag
W=<p_{1}, \dots , p_{3}>\oplus <u^{1}, \dots , u^{3}>\oplus <z_{1}, z_{2}>\oplus
<w_{1}, w_{2}>
\end{equation}

The space $<p_{1}, \dots , p_{3}>$ transforms according fundamental
representation $V$ of $GL(3)$; $<u^{1}, \dots , u^{3}>$ transforms by
contragradient representation $V^{\ast }$ of $GL(3)$ ; in both cases $SU_{L}(2)\times SU_{R}(2)$ factor acts trivially. The space $<z_{1}, z_{2}>$ is
irreducible two-dimensional representation $T_{L}$ of $SU_{L}(2)$ on which $GL(3)$ acts via $det$; $
<w_{1}, w_{2}>$ is two-dimensional representation $T_{R}$ of $SU_{R}(2)$ 
on which $GL(3)$ acts via $det^{-1}$. An algebra $\tilde{A}$ is defined as 
\begin{equation}\label{E:algA}
\tilde{A}={\mathbb C}[p_{1}, \dots , p_{3}, u^{1}, \dots, u^{3}, z_{1}, z_{2}, w_{1}, w_{2}]/I
\end{equation}
 where the ideal $I$ is generated by a single equation 
\begin{equation}\notag
p_{1}u^{1}+p_{2}u^{2}+p_{3}u^{3}
\end{equation}
An algebra $A$ is a subalgebra of $\tilde{A}$ generated by monomials of the
form $p_{\alpha }z_{i}\quad (1\leq \alpha \leq 3, 1\leq i\leq 2)$, $u^{\beta
}w_{j}\quad (1\leq \beta \leq 3, 1\leq j\leq 2)$
One can interpret algebra $A$
as a multigraded Serre algebra corresponding to a collection of
homogeneous line bundles over homogeneous space $F(1, 2)
\times{\mathbb CP}^1\times{\mathbb CP}^1$. We use notations
$F(1, 2)$ for the space of flags in in three dimensional complex
vector space. The symbol ${\mathbb CP}^1$ stands for one-
dimensional complex projective space.

Let $\Lambda \lbrack \pi _{\alpha i}]$ ($1\leq \alpha \leq 3, 1\leq i\leq 2$)
be a Grassmann algebra on $6$ anticommuting variables spanning $V\otimes
T_{L}$ representation of $GL(3)\otimes SU_{L}(2)$. A Grassmann algebra $\Lambda \lbrack \psi
_{j}^{\beta }]$ ($1\leq \beta \leq 3, 1\leq j\leq 2$) is generated by $
V^{\ast }\otimes T_{R}$. Introduce an algebra $\tilde{R}_{N=3}$
\begin{equation}\notag
\tilde{A}\otimes \Lambda \lbrack \pi _{\alpha i}]\otimes \Lambda \lbrack \psi
_{j}^{\beta }]
\end{equation}
The algebra $\tilde{R}_{N=3}$ carries a differential $d$, defined by the
formula 
\begin{align}
& d(\pi _{\alpha i})=p_{\alpha }z_{i}\quad (1\leq \alpha \leq 3, 1\leq i\leq
2) \\
& d(\psi _{j}^{\beta })=u^{\beta }w_{j}\quad (1\leq \beta \leq 3, 1\leq j\leq
2) \\
& d(p_{\alpha })=0 \\
& d(u^{\beta })=0 \\
& d(z_{i})=0 \\
& d(w_{j})=0
\end{align}

It is clear that the algebra 
\begin{equation}\label{E:algR}
R_{N=3}=A\otimes \Lambda[\pi_{\alpha i}]\otimes \Lambda[\psi^{\beta }_j]
\end{equation} is a differential graded subalgebra of $\tilde{R}_{N=3}$.

The claim is that the algebra $R_{N=3}$ is quasiisomorphic to $\Omega _{0}$
and $B_{0}.$ It is proved in the Appendix B.

\section{ A$_{\infty }$ -algebras and gauge theories.}\label{S:gauge}

 It is easy to formulate gauge theories in terms of $A_{\infty }-$
algebras.

\bigskip Let us consider for example 10D SUSY\ YM theory reduced to a point.
Corresponding action functional has the following form 
\begin{equation}\notag
S_{IKKT}(A, \chi)=tr(-\frac{1}{4}\delta^{ij}\delta ^{kl}[A_{i}, A_{k}][A_{j}, A_{l}]+\frac{1}{2}\Gamma _{\alpha\beta }^{i}[A_{i}, \chi ^{\alpha }]\chi ^{\beta })
\end{equation}
Here $A_{i}$ $i=1\dots 10$ is a set of even $N\times N$ matrices, $\chi
^{\alpha }$ , $\alpha =1\dots 16$ is a set of odd $N\times N$ matrices
\bigskip

The functional $S_{IKKT}$ is invariant with respect to infinitesimal gauge
transformations
\begin{equation}\notag
\delta A_{i}=[A_{i}, \varepsilon ]; \delta \chi ^{\alpha }=[\chi ^{\alpha
}, \varepsilon ]
\end{equation}
We can extend it to the solution of BV master equation in standard way 
\begin{equation}\notag
S=S_{IKKT}+trA^{\ast i}[A_{i}, c]+tr\chi _{\alpha }^{\ast }[\chi ^{\alpha
}, c]+\frac{1}{2}tr[c, c]c^{\ast }
\end{equation}
\bigskip Corresponding nilpotent vector field $Q$ can be written in the following
way
\begin{equation}\label{E:bv1}
QA^{\ast l}=\delta^{ij}\delta^{kl}[A_{i }, [A_{j}, A_{k}]]-\frac{1}{2}\Gamma _{\alpha \beta }^{l}\{\chi^{\alpha }, \chi ^{\beta }\}-[A^{\ast l}, c]
\end{equation}
\begin{equation}\label{E:bv2}
Q\chi _{\alpha }^{\ast}=-\Gamma _{\alpha \beta }^{i}[A_{i}, \chi ^{\beta }]-[\chi _{\alpha }^{\ast}, c]
\end{equation}
\begin{equation}\label{E:bv3}
Qc^{\ast }=-\lbrack A^{\ast i}, A_{i}]-[\chi _{\alpha }^{\ast }, \chi ^{\alpha}]+[c, c^{\ast }]
\end{equation}
\begin{equation}\label{E:bv4}
Qc=\frac{1}{2}\lbrack c, c]
\end{equation}
\begin{equation}\label{E:bv5}
QA_{ i}=[A_{i}, c]
\end{equation}
\begin{equation}\label{E:bv6}
Q\chi^{\alpha }=[\chi^{\alpha }, c]
\end{equation}
Equations of motion in BV-formalism are obtained as 
conditions of vanishing of RHS of (\ref{E:bv1}- \ref{E:bv6}). Using the 
connection between $Q$-manifolds and L$_{\infty}$- algebras described in Appendix C and considering Taylor series at a point $A_{ i}=\chi^{\alpha }=A^{\ast l}=\chi _{\alpha }^{\ast}=c^{\ast }=c=0$ we identify equations of motion with Maurer-Cartan equations for some $L_{\infty }-$ algebra. In our case this $
L_{\infty }$-algebra can be obtained from $A_{\infty }-$ algebra $\mathcal{A}
_{IKKT}$ by taking tensor product with matrix algebra and passing  to
corresponding $L_{\infty }-$ algebra. (This is true for equations of motion
in any gauge theory.) Regarding $A_{k}, \chi ^{\alpha }, c, A^{\ast
k}, \chi _{\alpha }^{\ast }, c^{\ast }$ as formal noncommutative variables we
can interpret (\ref{E:bv1}, \ref{E:bv2}, \ref{E:bv3}, \ref{E:bv4}, \ref{E:bv5}, \ref{E:bv6}) as a definition of the algebra $\mathcal{A}_{IKKT}$. More precisely the algebra $\mathcal{A}_{IKKT}$ can be considered as vector space spanned
by $A_{k}, \chi ^{\alpha }, c, A^{\ast k}, \chi _{\alpha }^{\ast }, c^{\ast }$
with operations $m_{2}$ (multiplication), $m_{3}$(Massey product) defined by the following formulas:
\begin{align}
&m_2(\chi ^{\alpha }, \chi ^{\beta })=\Gamma^{\alpha\beta}_k A^{\ast k}\\
&m_2(\chi ^{\alpha }, A_{k})=m_2(A_{k}, \chi ^{\alpha })=
\Gamma^{\alpha\beta}_k\chi _{\beta }^{\ast }\\
&m_2(\chi ^{\alpha }, \chi _{\beta }^{\ast })=
m_2(\chi _{\beta }^{\ast }, \chi ^{\alpha })=c^{\ast }\\
&m_2(A_{k}, A^{\ast k})=m_2(A^{\ast k}, A_{k})=c^{\ast }\\
&m_3(A_{k}, A_{l}, A_{m})=\delta_{kl}A^{\ast m}-\delta_{km}A^{\ast l}\\
&m_2(c, \bullet)=m_2(\bullet, c)= \bullet \\
\end{align}
All other products are equal to zero. 
$c$ is the unit of the A$_{\infty }-$algebra (for all $n$ -ary products 
(
$n \geq 3$) if at least one entry equal to the unit 
the whole product vanishes). 

All operations $m_{k}$ with $k\neq 2, 3$ vanish.

One can prove that $\mathcal{A}_{IKKT}$ is quasiisomorphic to differential
algebras $\Omega _{0}, B_{0}, $ etc described in (Sec. \ref{S:CH}); taking into
account that $m_{1}=0$ one can say that $\mathcal{A}_{IKKT}$ is a minimal
model of these algebras. The proof can be based on diagram techniques for
construction of minimal model developed in 
\cite{Merculov}, \cite{KS} and reviewed in Appendix C. (One can use, 
for example, the embedding of homology of $B_{0}$ into $B_{0}$
constructed in \cite{Berk3}.). 
Another proof will be given in 
\cite{MSch3}.

Notice that the higher multiplications ($k >2$) in 
A$_{\infty }-$ algebra come from quartic and higher terms 
in the action functional. It is easy to represent the 
action functional $S_{IKKT}$ in equivalent form containing 
at most 
cubic terms. (One should introduce the field strength
$F_{ij}=[A_i, A_j]$ as an independent field.) This remark
allows us to describe a smallest possible differential graded 
algebra quasiisomorphic to $\mathcal{A}_{IKKT}$.

Suppose $V$ is a ten- dimensional vector representation of 
$Spin(10)$, $S$ is a spinor representation, $S^*$ is a 
dual spinor representation. The algebra $C^{\bullet}$ has the following 
graded components:
$C_0={\mathbb C}$, $C_1=C_2=\{0\}$, $C_3=V$, $C_4=S$, $C_5=\Lambda^2(V)$, $C_6=\Lambda^2(V)$, $C_7=S^*$, $C_8=V$, $C_9=C_{10}=\{0\}$, $C_{11}={\mathbb C}$. All multiplication maps in this algebra are $Spin(10)$ equivariant as well as the differential. The algebra is graded commutative.

The differential is equal to zero on all components except $C_5$, where $d:C_5=\Lambda^2(V)\rightarrow \Lambda^2(V)=C_6$ is an isomorphism.

The space $C_0$ is generated by the unit. The multiplication defines a canonical pairing $C_i\otimes C_{11-i}\rightarrow C_{11}$, so we are dealing with a Frobenius algebra. Nontrivial multiplications are:
\begin{equation}\notag
C_3\otimes C_{3}=V\otimes V\rightarrow \Lambda^2(V)=C_6
\end{equation}
is the canonical projection, 
\begin{equation}\notag
C_3\otimes C_{4}=V\otimes S \overset{\Gamma}{\rightarrow} S^*=C_7
\end{equation}
is specified by means of gamma matrices. The map.
\begin{equation}\notag
C_5\otimes C_{4}=\Lambda^2(V)\otimes V \rightarrow V=C_8
\end{equation}
 is adjoint to inclusion $\Lambda^2(V)\rightarrow V\otimes V$.

One 
can extend the functional tr: $C_{11}\rightarrow {\mathbb C}$
to the map $C_{11}\otimes Mat_n \rightarrow 
{\mathbb C}$ using standard trace on algebra $Mat_n$;  we keep the 
notation $tr$ for extended functional. 

Consider a tensor product $C^{\bullet}\otimes Mat_n$.
The action functional corresponding to this algebra has the 
form 
\begin{equation}\label{E:eno}
S(a)=tr(\frac{1}{2}a*d(a)+\frac{2}{3}a*a*a)
\end{equation}
 for a field $a \in C^{\bullet}\otimes Mat_n$. 

It is easy to check that this action functional is equivalent to the 
cubic form of $S_{IKKT}$ .

\section{SYM algebra and Koszul duality.}

The algebra $F\left(\tilde{{\cal Q}}\right)$ of polynomials on the manifold
of pure spinors can be considered as a graded algebra with generators \ \ $u^{1}, \ldots , u^{n}$ \ and quadratic relations 
\begin{equation}\notag
u^{\alpha }\Gamma _{\alpha \beta }^{m}u^{\beta }=0, \quad u^{\alpha }u^{\beta
}=u^{\beta }u^{\alpha }.
\end{equation}
This fact permits us to study this algebra using well developed theory of
quadratic algebras. \ Recall, that a graded algebra \ $A=\Sigma A_{n}$ \ is
called a quadratic algebra if \ $A_{0}=\mathbb{C}, $ \ \ $W=A_{1}$ \
generates \ $A$ \ and all relations follow from quadratic relations \ $\sum\limits_{i, j}r_{ij}^{k}x^{i}x^{j}=0$ \ where \ $x^{1}, \ldots , x^{\dim W}
$ \ is a basis of \ $W=A_{1}.$ \ The space of quadratic relations (the
subspace of \ $W\otimes W$ \ spanned by \ $r_{ij}^{1}, r_{ij}^{2}, \ldots $)
\ will be denoted by \ $R.$ \ We can say that \ $A_{2}=W\otimes W/R$ \ and \ 
$A$ \ is a quotient of free algebra (tensor algebra) \ $\Sigma W^{n}$ \ with
respect to the ideal generated by \ $R.$ \ The dual quadratic algebra \ $%
A^{!}$ \ is defined as a free algebra \ $\Sigma \left(W^{\ast }\right) ^{\otimes n}$
\ generated by \ $W^{\ast }$ \ factorized with respect to the ideal
generated by \ $R^{\perp }\subset W^{\ast }\otimes W^{\ast }$ \ (here \ $%
R^{\perp }$ \ stands for the subspace of \ $W^{\ast }\otimes W^{\ast
}=\left(W\otimes W\right) ^{\ast }$ \ that is orthogonal to \ $R\subset
W\otimes W$). 
An important class of quadratic algebras consists of so called Koszul
algebras. For definition of this notion and for more details 
about duality of quadratic algebras see \cite{Priddy}, \cite{rqwerqe}, 
\cite{Backelin} or \cite{Bezr}.

The space of generators of \ $A=F\left(\tilde{{\cal Q}}\right) $ \ can be identified with the space of spinor representation \ $S; $ \ the space \ $R$
\ of relations is spanned by antisymmetric tensors and by ten-dimensional
subrepresentation \ $V\subset \left(S\otimes S\right) ^{sym}.$ \ The space
of generators of \ $A^{!}$ \ can be identified with \ $S^{\ast }$ \ (dual
spinors = right spinors); then \ $R^{\perp }$ \ is a subspace in \ $\left(
S^{\ast }\otimes S^{\ast }\right) ^{sym}$ \ orthogonal to \ $V$ \ (it is
generated by tensor squares of pure dual spinors). \ This means that \ $A^{!}
$ \ is generated by \ $\lambda \in S^{\ast }$ \ with relations 
\begin{equation}\label{E:lgtre}
\Gamma _{m_{1}, m_{2}, m_{3}, m_{4}, m_{5}}^{\alpha \beta }\left(\lambda
_{\alpha }\lambda _{\beta }+\lambda _{\beta }\lambda _{\alpha }\right) =0.
\end{equation}
where \ $\Gamma^{\alpha \beta} _{m_{1}, \ldots , m_{5}}$ \ stands for the skew-symmetrized product
of \ five \ $\Gamma -$matrices
\begin{equation}\label{E:kgfhdf}
\Gamma _{m_{1}}^{\alpha \delta_1}\Gamma _{m_{2} \delta_1 \delta_2}\Gamma _{m_{3}}^{\delta_2 \delta_3}\Gamma _{m_{4} \delta_3 \delta_4}\Gamma _{m_{5}}^{\delta_4 \beta}
\end{equation}

Notice, that these relations can be considered as conditions on (super)
commutator \ $\left[ \lambda _{\alpha }, \lambda _{\beta }\right]
_{+}=\lambda _{\alpha }\lambda _{\beta }+\lambda _{\beta }\lambda _{\alpha }.
$ \ Hence we can regard (\ref{E:lgtre}) as defining relations of a (super) Lie
algebra \ $L$ \ and the algebra \ $A^{!}=F\left(\tilde{{\cal Q}}\right)
^{!}$ \ can be regarded as its enveloping algebra: \ $F\left(
\tilde{{\cal Q}}%
\right) ^{!}=U\left(L\right) .$

The construction of the algebra $A=F\left(\tilde{{\cal Q}}\right)$ 
is a particular case of the following general construction. 
Let us consider a holomorphic line bundle $\alpha$ over 
complex manifold $M$ and the space $\Gamma_n$ 
 of holomorphic sections 
of the tensor power $\alpha^{\otimes n}$ of the bundle $\alpha$.
 The natural multiplication 
$\Gamma_m\otimes\Gamma_n \rightarrow 
\Gamma_{m+n}$ defines a structure of commutative 
associative graded algebra on the direct sum 
$\Gamma=\bigoplus_{n \geq 0} \Gamma_n$; this algebra is called 
Serre algebra corresponding to the pair $(M, \alpha)$. It is 
easy to generalize this construction to the case when we 
have a 
collection of line bundles $\alpha_1, \dots, \alpha_k$. In this case the direct sum of spaces $\Gamma_{n_1, \dots, n_k}$ of sections of line bundles $\alpha^{\otimes n_1}_1\otimes \dots \otimes \alpha^{\otimes n_k}_k$ ($n_1, \dots, n_k \geq 0$) carries a structure of commutative algebra which is called a multigraded Serre algebra. When $M=G/P$ is a compact complex homogeneous space of semisimple Lie group $G$ and $\alpha$ is a homogeneous line bundle one can prove that corresponding algebra is quadratic and that this algebra is Koszul algebra \cite{Bezr}. The algebra $A=F\left(\tilde{{\cal Q}}\right)$ can be obtained as a Serre algebra corresponding to $SO(10)/U(5)$, hence it is Koszul algebra.

Now we can consider the dual algebra to the algebra \ $B.$ \ The algebra \ $%
B $ \ is also a quadratic algebra with generators $u^{\alpha }, \theta
^{\alpha }$ \ and relation 
\begin{equation}\notag
u^{\alpha }\Gamma _{\alpha \beta }^{m}u^{\beta }=0, \ u^{\alpha }u^{\beta
}=u^{\beta }u^{\alpha}, \ \theta ^{\alpha }\theta ^{\beta }+\theta ^{\beta }\theta
^{\alpha }=0,  u^{\alpha }\theta^{\beta}=\theta^{\beta }u^{\alpha}
\end{equation}
Its dual algebra \ $B^{!}$ \ is an algebra with generators \ $\lambda
_{\alpha }, t_{\alpha }$ \ and relations \ 
\begin{equation}\notag
\Gamma _{m_{1}, \ldots m_{5}}^{\alpha \beta }\left(\lambda _{\alpha }\lambda
_{\beta }+\lambda _{\beta }\lambda _{\alpha }\right) =0, \ t_{\alpha
}t_{\beta }-t_{\beta }t_{\alpha }=0, \ \lambda _{\alpha }t_{\beta }-\lambda
_{\beta }t_{\alpha }=0.
\end{equation}
Again these relations involve only (anti)commutators and can be considered
as defining relation of (super) Lie algebra \ $\mathbb{L}.$ \ The algebra \ $%
B^{!}$ \ can be regarded as an enveloping algebra of \ $\mathbb{L}.$ \ The
differential acting on \ $B$ \ is defined by formulas \ 
$d\theta ^{\alpha }=u^{\alpha }, \
du^{\alpha }=0.$ \ The dual differential acting on \ $B^{!}$ \ obeys \ $%
d\lambda _{\alpha }=t_{\alpha }, \ dt_{\alpha }=0.$ \ 
It is clear from this
description that 
\begin{equation}\notag
A_{m}=\Gamma _{m}^{\alpha \beta }\lambda _{\alpha }\lambda _{\beta
}\text{ \ and \ }\chi^{\alpha }=
\Gamma _{m}^{\alpha \beta }\left[
\lambda _{\beta }, A^{m}\right] 
\end{equation}
satisfy \ $d A_{m}=0, $ \ $d\chi^{\alpha }=0.$ 

\ Let us denote
by SYM the subalgebra of the algebra \ $B^{!}$ \ generated by \ $A_{m}$ \ and \ $\chi^{\alpha }.$

It is easy to check that generators of SYM obey  relations 
\begin{equation}\label{E:SYM1}
[A_{i }, [A_{i}, A_{k}]]-\frac{1}{2}\Gamma _{\alpha \beta }^{k}\{\chi^{\alpha }, \chi ^{\beta }\}=0
\end{equation}
\begin{equation}\label{E:SYM2}
-\Gamma _{\alpha \beta }^{i}[A_{i}, \chi ^{\beta }]=0
\end{equation}
that follow from (\ref{E:lgtre}). 
 Comparing \ (\ref{E:SYM1}, \ref{E:SYM2}) \ with equations of motion of 10D SUSY\ YM reduced to a point we
obtain that the solutions to these equations of motion in \ $n\times n$ \
matrices can be identified with \ $n-$dimensional representations of the
algebra SYM.

It is easy to check that \ $SYM\subset Ker d\subset B_0^{!}.$ \ This inclusion determines a homomorphism of SYM into homology \ $H\left(B_0^{!}\right) $ \ of \ $B_0^{!}; $ \ one can prove that this homomorphism is an isomorphism (see \cite{MSch3}). \ This permits us to identify SYM with \ $H\left(B_0^{!}\right) $ \ and to say that the embedding of SYM into \ $B_0^{!}$ \ is a 
quasiisomorphism.

Notice, that an analog of algebra SYM for non-supersymmetric 
Yang-Mills theory (the Yang-Mills algebra YM) was considered in 
\cite{Nekrasov} and 
\cite{Connes-DV}; 
the paper \cite{Connes-DV} contains, in particular, the
construction of Koszul dual to YM.

\section{Appendix A. \\ Quasiisomorphism of Dolbeault and Berkovits algebras.}

Let us sketch the proof of quasiisomorphism of algebras $(\Omega _{0}, 
\overline{\partial })$ and $(B_{0}, d); $ the proof of quasiisomorphism
between $(\Omega , \overline{\partial })$ and $(B, d)$ is similar (see \cite{M6} for
more details).

Let us start with the analysis of the algebra $F(\widetilde{{\cal Q}})$ of
polynomial functions on the manifold of pure spinors $\widetilde{{\cal Q}}.$
Elements of $F(\widetilde{{\cal Q}})$ can be characterized as holomorphic functions on 
$\widetilde{{\cal Q}}$ that have polynomial growth at infinity and are bounded on
bounded subsets of $\widetilde{{\cal Q}}.$ We can work with the manifold $\widehat{{\cal Q}
}$ instead of $\widetilde{{\cal Q}}.$ (Recall that $\widetilde{{\cal Q}}$ is a total space
of principal $\mathbb{C}^{\ast }$-bundle $\alpha $ over ${\cal Q}$ and $\widehat{{\cal Q}}$
is a total space of line bundle $\widehat{\alpha }$ that is associate to
this $\mathbb{C}^{\ast }$-bundle.) Then we can characterize $F(\widetilde{{\cal Q}}
) $ as the space of holomorphic functions on $\widehat{{\cal Q}}$ having polynomial
growth at infinity. (Every continuous function of $\widehat{{\cal Q}}$ determines a
function on $\widetilde{{\cal Q}}$ that is bounded on every bounded domain. This
follows from corresponding fact about $\mathbb{C}$ and $\mathbb{C}^{\ast }.)$

Let us denote by $(\Omega (\widehat{{\cal Q}}), \overline{\partial })$ the algebra
of $(0, k)$-forms on $\widehat{{\cal Q}}$ having at most polynomial growth at
infinity. This algebra is quasiisomorphic to $F(\widetilde{{\cal Q}}); $ more
precisely the natural embedding of $F(\widetilde{{\cal Q}})$ into $\Omega (\widehat{
Q})$ is a $Spin(10, \mathbb{C})$-equivariant quasiisomorphism. We identified
already $H^{0}(\Omega (\widehat{{\cal Q}}), \overline{\partial })$ with $F(
\widetilde{{\cal Q}}); $ it remains to verify that $H^{k}(\Omega (\widehat{{\cal Q}}), 
\overline{\partial })=H^{0, k}(\widehat{{\cal Q}})=0$ for $k>0.$ The calculation of $
H^{0, k}(\widehat{{\cal Q}})$ can be reduced to the calculation of Dolbeault
cohomology $H^{0, k}(\widehat{\beta }^{\otimes n})$ $n \geq 0$ of
 line bundles $\widehat{\beta }^{\otimes n}$over ${\cal Q}$. The 
line bundle $\widehat{\beta }$ is dual to $\widehat{\alpha }$; 
these cohomology vanish unless $k=0$.

Using the quasiisomorphism between $(\Omega (\widehat{{\cal Q}}), \overline{\partial 
})$ and $F$($\widetilde{{\cal Q}})$ we obtain that the natural embedding of $
B_{0}=(F(\widetilde{{\cal Q}})\otimes \Lambda (S^{\ast }), d)$ into the algebra $($ $
\Omega (\widehat{{\cal Q}})\otimes \Lambda (S^{\ast }), \overline{\partial }+d)$ is
a quasiisomorphism. (To check that the latter algebra has the same
cohomology as $B_{0}$ one can consider it as a bicomplex with differentials $
\overline{\partial }$ and $d$ and prove that the corresponding spectral
sequence degenerates, hence the cohomology is equal $H(H(\overline{\partial }
), \widetilde{d})$ where $H(\overline{\partial })$ stands for the cohomology
of $\Omega (\widehat{{\cal Q}})\otimes \Lambda (S^{\ast })$ with respect to $
\overline{\partial }$ and $\widetilde{d}$ denotes the differential on $H(
\overline{\partial })$ induced by $d$.)

\bigskip The algebra $($ $\Omega (\widehat{{\cal Q}})\otimes \Lambda (S^{\ast }), 
\overline{\partial }+d)$ is quasiisomorphic to the algebra $\Omega (\widehat{
{\cal Q}}\times \Pi S)$ of $(0, k)$-forms on the manifold $\widehat{{\cal Q}}\times \Pi S$
with differential $\overline{\partial }+d.$ (This follows from decomposition 
$\Omega (\widehat{{\cal Q}}\times \Pi S)=$ $\Omega (\widehat{{\cal Q}})\otimes \Lambda
(S^{\ast })\otimes \Lambda (\overline{S}^{\ast })\otimes S (\overline{S}^{\ast })$. The natural projection $
\widehat{{\cal Q}}\times \Pi S$ onto ${\cal R}_{0}$ induces a homomorphism of $(\Omega
_{0}, \overline{\partial })$ into $($ $\Omega (\widehat{{\cal Q}}\times \Pi S), 
\overline{\partial }+d); $ one can check that this homomorphism is a
quasiisomorphism. To construct a map $\widehat{{\cal Q}}\times \Pi S\rightarrow
{\cal R}_{0}$ we interpret $\widehat{{\cal Q}}\times \Pi S$ as a total space of a bundle
over ${\cal Q}=SO(10, \mathbb{R})/U(5)$ that corresponds to representation $\lambda $
of $U(5)$ on $\mathbb{C}\times \Pi S; $ it is easy to check that $\lambda $
can be expressed in terms of vector representation $W$ of $U(5)$ as 
\begin{equation}\label{E:jkl}
\lambda =(\det W)^{-1/2}, ({\mathbb C}+\Lambda ^{2}(W)+\Lambda ^{4}(W))\otimes (\det W)^{-1/2})
\end{equation}
Denote a generator of linear space $(\det W)^{-1/2}, \{0\})\subset 
\lambda$ by $\xi$ , the generator of $(\{0\}, det W^{-1/2}+\{0\})\subset \lambda $ by $x$.
As we have seen ${\cal R}_{0}$ can be interpreted in similar way by 
means
of representation 
\begin{equation}\notag
\varkappa =(\Lambda ^{2}(W)+\Lambda ^{4}(W))\otimes (\det W)^{-1/2}
\end{equation}
The obvious intertwiner between $\lambda $ and $\varkappa $ gives rise to
the map we need. This map induces a homomorphism of $\Omega _{0}=\Omega
({\cal R}_{0})$ into $\Omega (\widehat{{\cal Q}}\times \Pi S)$.
To prove that this homomorphism is a quasiisomorphism we notice 
that in coordinates (\ref{E:jkl}) the differential $d$ is given 
by the formula $x\frac{\partial}{\partial \xi}+\bar{x}\frac{\partial
}{\partial \bar{ \xi} }$. It means that locally $x$ and $ \xi$ 
($ \bar{x}$ and $ \bar{ \xi}$)are contractible pairs and locally 
over $\cal Q$ the inclusion $i$ is quasiisomorphism. One can globalize 
 this statement using a partition of unity.

\section{Appendix B. \\ Quasiisomorphic Koszul complexes.}

Let us consider a differential (super)commutative associative algebra $
\mathcal{K}$ and $n$ even elements 
$f_{1}, ..., f_{n}\in \mathcal{K}$ 
obeying $df_{1}=...=df_{n}=0.$ Then we can define differential algebra $\mathcal{K}
^{f_{1}, ..., f_{n}}$ (Koszul complex) adding ''ghosts'' $c_{1}, ..., c_{n}$ and
-extending the differential by the formula $dc_{i}=f_{i}.$ (More formally, $
\mathcal{K}^{f_{1}, ..., f_{n}}$ is defined as a tensor product of $\mathcal{K}
$ with exterior algebra generated by $c_{1}, ..., c_{n}.$) Notice that the
algebra $\mathcal{K}^{f_{1}, ..., f_{n}}$ does not change if we replace $
f_{1}, ..., f_{n}$ with their linear combinations spanning the same ideal. One
says that $f_{1}, ..., f_{n}$ is a regular sequence if for every $r=0, ..., n-1$
the elements $f_{r+1}, ..., f_{n}$ do not divide zero in $\mathcal{K}
/(f_{1}, ..., f_{r}).$ Here $(f_{1}, ..., f_{r})$ stands for the ideal generated
by $f_{1}, ..., f_{r}.$ If the sequence $f_{1}, ..., f_{n}$ is regular then the
natural homomorphism $\mathcal{K}^{f_{1}, ..., f_{n}}$ into $\mathcal{K}
/(f_{1}, ..., f_{r})$ is a quasiisomorphism. (This statement is well known for
the case when the differential in $\mathcal{K}$ is trivial. Then $\mathcal{K}
^{f_{1}, ..., f_{n}}$ is called Koszul resolution of $\mathcal{K}
/(f_{1}, ..., f_{r}).$ The generalization to the case of non-trivial
differential in $\mathcal{K}$ is obvious.)

Let us assume now that the sequence $f_{1}, ..., f_{n}$ is not regular, but
its subsequence $f_{r+1}, ..., f_{n}$ is regular. Then one can prove that the
algebra $\mathcal{K}^{f_{1}, ..., f_{n}}$ is quasiisomorphic to the algebra $
\mathcal{K}^{f_{1}, ..., f_{r}}/(f_{r+1}, ..., f_{n}).$ This follows immediately
from the remark that 
\begin{equation}\notag
\mathcal{K}^{f_{1}, ..., f_{n}}=(\mathcal{K}^{f_{1}, ..., , f_{r}})^{f_{r+1}, ..., f_{n}}.
\end{equation}

We can apply the above statement to construct algebras that are
quasiisomorphic to Berkovits algebra. We start with the algebra $\mathcal{K}=F(\widetilde{{\cal Q}})$ of polynomial functions on the space of pure spinors: $
\mathcal{K}=\mathbb{C}[u^{1}, ..., u^{16}]/(u\Gamma u).$ It is easy to check
that Berkovits algebra $(B_{0}, d)$ is isomorphic to $\mathcal{K}
^{u_{1}, ..., , u_{16}}.$ We can verify that $B_{0}$ is quasiisomorphic to the
algebra $R_{N=3}$ using the following construction: we notice that the Lie algebra ${\mathfrak gl}(3)\times{\mathfrak 
su}(2)_L \times 
{\mathfrak su}(2)_R $ is a subalgebra of ${\mathfrak so}(10)$. 
The spin representation spanned by $<u^{1}, ..., u^{16}>$ restricted 
on ${\mathfrak gl}(3)\times{\mathfrak su}(2)_L \times {\mathfrak 
su}(2)_R $ splits into $V \otimes T_L \oplus V^* \otimes T_R 
\oplus T_L \oplus T_R $ where $V$ is a fundamental 
representation of ${\mathfrak gl}(3)$,  $V^*$ is its dual,  
$T_L$ is a two-dimensional representation of 
${\mathfrak su}(2)_R$. 
The ten-dimensional representation of $Spin(10)$ decomposes as 
$V+V^*+T_L \otimes T_R$. It is useful to write down pure spinor equation in this context. The coordinates of spinor are $t_{\alpha i}, s_j^{\beta}, z_i, w_j$, where $1 \leq ij \leq 2$, $1\leq \alpha, \beta \leq 3$.
The equations are
\begin{align}
&t_{\alpha 1}z_2-t_{\alpha 2}z_1=det(s_j^{\beta})\label{e:pbw1}\\
&s^{\alpha}_ 1w_2-s^{\alpha}_2w_1=det(t_{\alpha i})\label{e:pbw2}\\
&s^{\alpha}_ it_{\alpha j}=0\label{e:pbw3}
\end{align}
Expressions $det(s_j^{\beta})$, $det(t_{\alpha i})$ stand for vector formed by principal minors of $2 \times 3$ matrices.

 It can be shown using Maple package Gr\"{o}bner that the 
algebra $\mathcal{K}$ is a free ${\mathbb C}[w_i, z_j]$ module, i.e. it 
is equal to ${\cal N}\otimes {\mathbb C}[w_i, z_j]$ for some linear space ${\cal N}$. 
It is well known(e.g. see \cite{McL}) that in the case of
polynomial algebra ${\cal C}= 
{\mathbb C}[x_1, \dots, x_n]$ the cohomology 
$H^{\bullet}({\cal C} ^{x_1, \dots, x_r})$ is equal to $\{0\}$ in all 
degrees but zero where it is equal to ${\cal C} /{(x_1, \dots, x_r)}$
. This implies 
that $H^{\bullet}(V\otimes A ^{x_1, \dots, x_r})=V\otimes H^{\bullet}(A ^{x_1, \dots, x_r})$.
For the algebra at hand this means that 
$H^{\bullet}(\mathcal{K}^{w_i, z_j})=H^{\bullet}(V\otimes{\mathbb C}[w_i, z_j] ^{w_i, z_j})=V \otimes H^{\bullet}({\mathbb C}[w_i, z_j] ^{w_i, z_j})$. Hence the complex $\mathcal{K}^{w_i, z_j}$ has cohomology only in zero degree.

Define a map $\rho:\mathcal{K}/_{(w_i, z_j)}\rightarrow \tilde A$ by the rule
\begin{align}
&\rho(t_{\alpha i})=p_{\alpha }z_i \notag \\
&\rho(s^{\beta }_j)=u^{\beta }w_j \notag
\end{align}
(recall that $\tilde A$ was defined in (Sec. \ref{S:CH}) by means of (Equ. \ref{E:algA}))
It is easy to check correctness of this map. Its image is a 
subalgebra $A \subset \tilde A $. 
 We interpreted the algebra $A$ as a Serre algebra. Using a theorem that 
a 
multigraded Serre algebra of a compact homogeneous space of a 
semisimple group is quadratic (see \cite{Bezr} and references 
therein) , it is not hard to show that the map $\rho$ is an 
isomorphism of $\mathcal{K}/_{(w_i, z_j)}$ and algebra 
$A$. This means that differential graded algebra $\mathcal{K}^{w_i, z_j}$ is quasiisomorphic to $A$ , hence $\mathcal{K}^{u_{1}, ..., , u_{16}}$ is quasiisomorphic to $R_{N=3}$.
D. 

Piontkovski pointed out to us that the length of 
a maximal regular sequence for algebras with duality in cohomology
 of the Koszul complex (so called Gorenstein algebras) is equal 
to the dimension of the underlying affine manifold. Every regular 
sequence can be extended to the maximal one. Maximal regular 
sequences form a dense set (in a suitable topology) among all 
sequences of the same length. 
For the algebra $F(\widetilde{{\cal Q}})$ the maximal length 
of regular sequence is equal to 11. The maximal regular
sequence leads to differential algebra closely related 
to the algebra $C^{\bullet}$ described in (Sec. \ref{S:gauge}).

\section{Appendix C. \\ L$_{\infty }$  and A$_{\infty }$ algebras }

The BV formalism is based on consideration of classical master
equation $\{S, S\}=0$ on odd symplectic manifold 
($P$-manifold). Every solution
of classical master equation generates an odd vector field $Q$ corresponding
to the first order differential operator $Q:\varphi \rightarrow \{\varphi
, S\}$. This remark permits us to say that a geometric counterpart of
solution of classical master equation is a $PQ$-manifold, 
i.e. a $P$-manifold X
equipped with an odd vector field $Q$ obeying $Q^{2}=0$ and $L_{Q}\sigma =0$
. Here $\sigma $ is an odd symplectic form on X and $L_{Q}$ is a Lie
derivative along vector field $Q$. A manifold equipped with an odd vector
field $Q$ obeying $Q^{2}=0$ is called a $Q$-manifold. One 
can say that
$PQ$-manifold is equipped with structures of $P$-manifold 
and $Q$-manifold and
these structures are compatible.

Let $R$ denote the zero locus of $Q$ on $Q$-manifold $X$. Then the
differential $Q_{x}$ of $Q$ at a point $x\in R$ can be considered as a
linear operator acting on a tangent space $T_{x}Q$. One can identify $KerQ_{x}$ with the tangent space $T_{x}R$, the spaces $ImQ_{x}$ specify a
foliation of $R$; we denote a space of leaves of this foliation by $\tilde{R}
$. If $Q$ corresponds to the solution $S$ of the classical master equation, 
the zeros of $Q$ coincide with extrema of $S$ (solutions to the classical
equations of motion) and $\tilde{R}$ can be interpreted as the moduli space of
solutions to the classical equations of motion.

There is some freedom in presenting classical mechanical system in BV-form.
This freedom leads to notion of equivalent (quasiisomorphic) $Q$-manifolds
and equivalent (quasiisomorphic)$PQ-$ manifolds. Let us suppose that $f$ is
a map of $Q$- manifold into $Q$- manifold $X^{\prime }$ 
that is compatible
with vector field $Q$ i.e. $f_{\ast} Q=Q$. (We use the same notation
for nilpotent vector field on $X$ and on $X^{\prime }.$) Then the zero locus 
$R$ of $Q$ on $X$ goes to zero locus $R^{\prime }$ of $Q$ on $X^{\prime }$
and $\tilde{R}$ is mapped into $\tilde{R}^{\prime }$. We say that $f$ is
quasiisomorphism between $X$ and $X^{\prime }$ if correspondence between $
\tilde{R}$ and $\tilde{R}^{\prime }$ is bijective.

The notion of $Q$-manifold is closely related to the notion 
of L$_{\infty }$
-algebra. Namely if $p$ is a point of $M$ where the vector field $Q$ vanishes
the coefficients of Taylor expansion of $Q$ at the point $m$ specify a
structure of L$_{\infty }$-algebra on the tangent space $E=\Pi 
T_{x_{0}}(M)$.
More precisely, we consider the (formal) Taylor series

\begin{equation}\label{E:ndj}
Q^{a}(z)=\sum\limits_{k=1}^{\infty }\sum\limits_{i_{1}, ..., i_{k}}{
^{(k)}m_{i_{1}, ..., i_{k}}^{a}}z^{i_{1}}...z^{i_{k}} 
\end{equation}
of the vector field $Q$ with respect to the local coordinate system $
z^{1}, ..., z^{N}$ centered in $p\in R$. Coefficients in (\ref{E:ndj}) are symmetric
in the sense of superalgebra, i.e. symmetric with respect to transpositions
of two even indexes or an even index with an odd one and antisymmetric with
respect to transpositions of two odd indexes (``parity of index $i$'' means
parity of the corresponding coordinate $z^{i}$). We assume that the field $Q$
is smooth.

It is easy to check that the condition $\{ Q, Q \} =0$ is equivalent to the
following relations for the coefficients $^{(k)}m^a_{i_1, ..., i_k}$:

\begin{equation} \label{E:asd}
\sum\limits^{n+1}_{\overset{k, l=0}{ k+l=n+1}} \sum\limits^{k}_{p=1} \sum_{
\overset{\overset{perm}{i_1, ..., \hat{\imath}_p, ..., i_k}}{ j_1, ..., j_l}} \pm {
^{(k)}m^a_{i_1, ..., i_p, ..., i_k}} {^{(l)}m^{i_p}_{j_1, ..., j_l}} = 0
\end{equation}
where $\pm$ depends on the particular permutation.

Let us write the first relations, assuming for simplicity that $^{(3)}m=0$ . It
is more convenient to use instead of $^{(k)}m$ defined above the following
objects: 
\begin{equation}\notag
d^a_b = {^{(1)}m^{a^{\prime}}_{b^{\prime}}}
\end{equation}
and 
\begin{equation}\notag
f^a_{bc} = \pm {^{(2)}m^{a^{\prime}}_{b^{\prime}c^{\prime}}}
\end{equation}
where 
\begin{equation}\notag
\pm = (-1)^{(\epsilon(a^{\prime})+1) \epsilon(b^{\prime})}
\end{equation}
Here $\epsilon(a)$ denotes the parity of the index $a$, and the parity of
the indexes $a, b$ and $c$ is opposite to the parity of the corresponding
indexes in the right hand side: $\epsilon(a)= \epsilon(a^{\prime})+1$, etc.
{}From these formulas one can easily get the symmetry condition for $
f^a_{bc} $:

\begin{equation} \label{E:pof}
f^a_{bc} = (-1)^{\epsilon(b) \epsilon(c) +1} f^a_{cb}
\end{equation}

Then we have 
\begin{equation} \label{E:bta}
d^m_b d^c_m = 0
\end{equation}
\begin{equation} \label{E:htr}
f^r_{mb} d^m_c + (-1)^{\epsilon(b)\epsilon(c)} f^r_{mc} d^m_b + d^r_m
f^m_{cb} = 0
\end{equation}

\begin{equation} \label{E:sdf}
f^r_{mb} f^m_{cd} +(-1)^{(\epsilon(b) + \epsilon(d))\epsilon(c)} f^r_{mc}
f^m_{db} +(-1)^{(\epsilon(c)+\epsilon(d))\epsilon(b)} f^r_{md} f^m_{bc} =0
\end{equation}

Relation (\ref{E:sdf}) together with relation (\ref{E:pof}) means that the $
f^a_{bc}$ can be considered as structure constants of a super Lie algebra.
The matrix $d^a_b$ determines an odd linear operator $d$ satisfying $d^2=0$
(this follows from (\ref{E:bta})). The relation (\ref{E:htr}) can be considered
as a compatibility condition of the Lie algebra structure and the
differential $d$.

In a more invariant way we can say that the coefficients $^{(k)}m$ determine
an odd linear map of $k^{th}$ symmetric tensor power of $S^k(T_pM)$ into $T_pM$. This map
induces a map $m_k:S^k(V) \rightarrow V$, where $V=\Pi T_p M$. Here 
$m_k$ is odd for odd $k$ and even for even $k$. The map $m_1$ determines
a differential in $V$ and $m_2$ determines a binary operation there. The
relations (\ref{E:pof}-\ref{E:sdf}) show that in the case when $^{(3)}m=0$ the
space $V$ has a structure of a differential Lie (super)algebra.

If $^{(3)}m\ne 0$ the Jacobi identity (\ref{E:sdf}) should be replaced with
the identity involving $^{(3)}m$ (the so called homotopy Jacobi identity).
However taking homology $H(V)$ with respect to the differential $m_1 =d$
we get a Lie algebra structure on $H(V)$.

One can consider $m_k$ as a $k$-ary operation on $V$. Relation (\ref{E:asd}) can be rewritten as a set of relations on the operations $m_k$. A
linear space provided with operations $m_k$ satisfying these relations is
called a $L_{\infty}$-algebra or strongly homotopy Lie algebra. (The name
 homotopy Lie algebra is used when there are only $m_1, m_2$
and $m_3$ satisfying the corresponding relations.) The notion of strong
homotopy algebra was introduced by J. Stasheff who realized 
also that this
algebraic structure appears in string field theory \cite{St}.

The construction above gives a structure of $L_{\infty}$-algebra to the
space $V =\Pi T_pM$ where $p$ is a stationary point of an odd vector field $
Q $ satisfying $\{ Q, Q \} =0$. It is possible also to 
include an
``operation'' $^{(0)}m$ in the definition of $L_{\infty}$-algebra; then the
structure of an $L_{\infty}$-algebra arises in the space $V=\Pi T_pM$ at
every point $p$ of the $Q$-manifold $M$. However we will not use this modified definition.

One can say that L$_{\infty }$-algebra is a formal 
$Q$-manifold. (The space of functions on formal 
$(m|n)$-dimensional supermanifold can be identified 
with supercommutative algebra $\hat{C}^{m|n}$ of formal series 
with respect to commuting variables $x^1, \dots, x^m$ and anticommuting 
variables $\zeta^1, \dots, \zeta^n$. The algebra $\hat{C}^{m|n}$ can be 
considered as completion of free nonunital 
supercommutative algebra ${C}^{m|n}$ generated by even elements 
$x^1, \dots, x^m$ and odd elements $\zeta^1, \dots, \zeta^n$. See \cite{25} 
for discussion of formal and partially formal supermanifolds.) 

Maurer-Cartan equation in L$_{\infty }$-algebra has the form

\begin{equation}\notag
\sum_{k=1}^{\infty}\frac{1}{k!}m_k(a, \dots, a)=0
\end{equation}

Interpreting L$_{\infty }$-algebra as a $Q$-manifold one 
can identify the set
of solutions to the Maurer-Cartan equation with 
zero locus $R$ of vector
field $Q.$ The moduli space of solutions to the Maurer-Cartan equation
is obtained from this set by means of some identification; it
corresponds to $\tilde{R}$.

To talk about solutions to Maurer-Cartan equation one should have a 
notion of a point of formal manifold; see \cite{25} for discussion of this 
notion. 

Notice that in the case when the vector field $Q$ is polynomial 
(i.e. we have only finite number of operations $m_k$) we can 
work with a free algebra $T_n$ instead of its completion $\hat{T}_n$.

In the case of graded L$_{\infty}$-algebra we can also work with the free algebra $T_n$. 

If $X$ is a $PQ$-manifold and the field $Q$ vanishes at 
$x_{0}$ we can say
that L$_{\infty }$-algebra constructed above is equipped with an odd
nondegenerate inner product. More precisely there exists such an odd
bilinear form $<., .>$ on $E$ that $<a_{0}, m_{k}(a_{1}, ..., a_{k})>=\mu_k
(a_{0}, a_{1}, ..., a_{k})$ is cyclically (graded)symmetric (to construct such
an inner product we notice that in a neighborhood of $x_{0}$ we can find
coordinates in a such a way that coefficients of symplectic form are
constant). We can identify an L$_{\infty }$-algebra algebra with a
nondegenerate inner product with a formal $PQ$-manifold. 
Using this
identification we can say that Maurer-Cartan equations are equations of
motion corresponding to the action functional

\begin{equation}\label{A:CS}
S(a)=\sum_{k=1}^{\infty}\frac{1}{k!}\mu_k(a, \dots, a)
\end{equation}

This statement is a particular case of general fact that the equation $Qx=0$
can be regarded as an equation of motion corresponding to BV-action
functional $S.$

The action (\ref{A:CS}) can be considered as generalization of Chern-Simons action. (If an L$_\infty$algebra is a differential Lie algebra, i.e. $\mu_k=0$ for $n\geq 3$, then (\ref{A:CS}) looks as standard Chern-Simons action. One can prove that any L$_\infty$algebra is quasiisomorphic to differential Lie algebra. This means that every action can be represented in Chern-Simons form.)

The notion of $Q$-manifold can be defined also for noncommutative spaces.
An associative $\mathbb{Z}_{2}$ graded algebra can be regarded
as an algebra of functions on noncommutative (super)space. Vector fields are
identified with derivations (infinitesimal automorphisms), odd vector fields
with odd (parity changing) derivations. Noncommutative $Q$-manifold is a differential 
associative $\mathbb{Z}_{2}$ graded algebra, i.e an algebra equipped with an odd (parity
reversing) derivation $Q$ obeying $Q^{2}=0$.

One can define an A$_{\infty }$-algebra as a formal non-commutative $Q$
-manifold. A formal non-commutative manifold for us will mean  a
manifold with coordinates $x^{1}, ..., x^{n}$ that do not satisfy any
relations. In other words the algebra of functions on such manifold is a
completion $\widehat{T}_{n}$ of free algebra $T_{n}$ generated by $
x^{1}, ..., x^{n}$. More precisely, $\widehat{T}_{n}$ consists of all
infinite series in noncommuting variables $x^{1}, ..., x^{n}$.
We assume that coordinates $x^{1}, ..., x^{n}$ and hence the algebra
they generate $A$ are $\mathbb{Z}_{2}$- graded (i.e. some 
of the coordinates
are considered as even and some as odd). Notice that we did not assume that the algebra $T_{n}$ is unital; in other word a series
belonging to $\widehat{T}_{n}$ cannot contain a constant term. A vector
field is identified with a derivation (infinitesimal automorphism) of this
algebra. A vector field $Q$ specifying an $
A_{\infty }$-algebra structure is an odd derivation obeying $Q^{2}=0$. 
It is sufficient
to specify $Q$ on generators $x^{1}, ..., x^{n}$: 
\begin{equation}\notag
Qx^{i}=\sum m_{i_{1}, ..., i_{k}}^{i}x^{i_{1}}...x^{i_{n}}
\end{equation}

The vector field $Q$ is uniquely determined by its ``Taylor 
coefficients''
$m_{i_{1}, ..., i_{k}}^{i}$. We consider them as polylinear operations defined by the formula
\begin{equation}\label{E:oper}
m_{k}(e_{i_{1}}, ..., e_{i_{k}})=\pm m_{i_{1}, ..., i_{k}}^{a}e_{a}
\end{equation}
 We assume that elements $e_{1}, ..., e_{n}$ are in 1-1
correspondence with $x^{1}, ..., x^{n}$ but the parity of $e_{i}$ is opposite
to parity of $x^{i}$.
If $m_k=0$ for $k\geq3$ then its follows from the above relations that $m_1, m_2$ specify of differential associative algebra. Conversely, every differential associative algebra can be considered as A$_{\infty}$-algebra; the space $\hat{T}$ is dual to the space of Hochschild cochains and $Q$ is induced by Hochschild differential.

 The condition $Q^2=0$ can be rewritten
as a sequence of quadratic equations

$$\sum_{i+j=n+1}\sum_{0\le l\le i}\epsilon(l, j)
m_i(a_0, ..., a_{l-1}, m_j(a_l, ..., a_{l+j-1}), a_{l+j}, ..., a_n)=0$$
where $a_m \in A$, and $\epsilon(l, j)=
(-1)^{j\sum_{0\le s\le l-1}deg(a_s)+l(j-1)+j(i-1)}$.
In particular, $m_1^2=0$.
\bigskip

In coordinate-free language we start with 
${\mathbb Z}_2$-graded vector space $V$ 
and define tensor algebra $\hat{T}(V)$ as 
completion of tensor algebra 
$${T}(V)=\bigoplus_{n \geq 0} V^{\otimes n}$$ 
We consider $\hat{T}(V)$ 
as topological ${\mathbb Z}_2$-graded algebra. 
A continuous derivation $Q$ of $\hat{T}(V)$ is completely 
determined by its values on $V^{\otimes 1}\subset \hat{T}(V)$. 
In other 
words to specify $Q$ one needs to fix linear maps 
$\mu_k:V\rightarrow V^{\otimes k}$. These maps determine 
${\mathbb C}$-linear operations $m_k:A^{\otimes k}\rightarrow A$, 
where $A$ stands for the space obtained from $V^*$ by means of 
parity reversion: $A=\Pi V^*$. If $Q^2=0$, then the operations 
specify a structure of A$_{\infty}$-algebra on $A$. Notice that since 
$Q$ is 
odd,  the parity of $m_k(a_1, \dots, a_k)$ is equal to 
$k(\sum deg a_i)$ mod ${\mathbb Z}_2$, where $deg a_i$ stands for the parity of $a_i$

An element $1$ is called the unit of A$_{\infty}$-algebra $A$ if it is
the unit of binary operation $m_2$ and in all higher order operations
we have $m_k(\dots, 1, \dots)=0$.
 
It is convenient to assume that $A$ has no unit or that it is obtained
from nonunital algebra by means of adjunction of a unit.

Considering L$_{\infty}$-algebras as formal $Q$- manifolds 
we can define an L$_{\infty}$-morphism of 
L$_{\infty}$-algebras as 
 a map of $Q$- manifolds which is compatible with $Q$ .
(One says that that a map $f: M \rightarrow M'$ of $Q$- 
manifolds is compatible with $Q$ if $f_{\ast} Q=Q'$. We use
here the same notation $Q$ for vector fields specifying
$Q$-structure on $M$ and $M'$.)
 As in the case of definition of structure maps $m_k$ in 
(\ref{E:oper}) of L$_{\infty}$- algebra the components 
of L$_{\infty}$-morphism can be read off from Taylor 
coefficients of the map $f$. This means that an 
L$_{\infty}$-morphism from $L$ to $L'$ is a system of 
linear transformation $f_k:L^{\otimes k}\rightarrow L'$ 
obeying nonlinear equations that follow from compatibility
between $f$ and $Q$.

It is easy to check that $f_1m_1=m'_1f_1$. This means that 
L$_{\infty}$-morphism induces a homomorphism of homology 
$H(L, m_1)\rightarrow H(L', m'_1)$. One says that 
L$_{\infty}$-morphism morphism is a quasiisomorphism if it 
induces an isomorphism on homology. All important notions
are invariant with respect to quasiisomorphism; in particular, 
the moduli space of solutions to Maurer-Cartan equation does not
change if we replace an algebra with quasiisomorphic algebra.

An L$_{\infty}$-algebra $L$ is called minimal if $m_1=0$. 
One can prove that every L$_{\infty}$-algebra is 
quasiisomorphic to a minimal L$_{\infty}$-algebra. 
(In other words, there exists a minimal L$_{\infty}$- 
structure on 
$H(L, m_1)$ and a quasiisomorphism between $L$ and 
$H(L, m_1)$ with this L$_{\infty}$-structure.) 
More precisely, one can say that every L$_{\infty}$-algebra
is isomorphic to a direct sum of minimal algebra and trivial
algebra. (We say that L$_{\infty}$-algebra is trivial if
it has only one non-zero operation $m_1$ and trivial homology.)

Two minimal L$_{\infty}$-algebras are quasiisomorphic 
iff 
they are isomorphic 
(i.e. they are connected by invertible 
L$_{\infty}$-morphism 
).
On the other hand every  L$_{\infty}$-algebra is 
quasiisomorphic to differential Lie algebra 
(i.e. to L$_{\infty}$-algebras algebra with $m_k=0$ for $k\geq 3 $). 
In other words quasiisomorphism (L$_{\infty}$-quasiisomorphism) classes 
of differential Lie algebras can be identified with quasiisomorphism 
classes of L$_{\infty}$-algebras and isomorphism classes of 
minimal L$_{\infty}$-algebras.

Similar statements are true for A$_{\infty}$-algebras and differential 
associative algebras if we modify definitions in appropriate way.(For
example, the definition of A$_{\infty}$-morphism is analogous to the
definition of L$_{\infty}$-morphism.)

The above statements give us an explanation of the role of 
A$_{\infty}$-algebras. 
Differential associative algebras are ubiquitous 
in physics. In particular, BRST operator can be considered as a  
differential on an appropriate algebra of operators. Usually two 
quasiisomorphic algebras are physically equivalent. This fact allows 
us to replace a differential algebra defined by BRST operator with much 
smaller A$_{\infty}$-algebra on the space of observables (on 
BRST-cohomology). From the other side one can avoid the use of 
A$_{\infty}$-algebras working with larger differential associative 
algebras. 

 There exists a diagram technique, that permits us to calculate explicitly
the minimal model of A$_{\infty}$-algebra or L$_{\infty}$-algebra.
More generally, let us denote by $B$ an $m_1$-invariant ${\mathbb 
Z}_2$-graded subspace of a nonunital 
A$_{\infty}$-algebra $A$. Let us assume that there exist a linear 
operator
$P$:$A \rightarrow A$ obeying $P^2=P$ that projects $A$ onto $B$ 
and 
commutes with $m_1$. We assume that the projection $P$ is 
homotopic to the identity map, i.e. there exists
an odd operator $H$ obeying $1-P=m_1 H + H m_1$. In these 
assumptions the embedding $i:B \rightarrow A$ and
the projection $p: A \rightarrow B $ induce 
isomorphisms between homology of $B$ and $A$. (The projection
$p$ is defined by the formula $P=i \circ p$.)
Following \cite{KS} we define a structure of
 A$_{\infty}$-algebra on $B$. If the differential $m_1$ acts
trivially on $B$ this construction gives a minimal model of $A$.

One can introduce a sequence of linear operations 
$m_n^B:B^{\otimes n}\to \Pi^nB$ in the following way

a) $m_1^B:=d^B=p\circ m_1\circ i$; 

b) $m_2^B=p\circ m_2\circ (i\otimes i)$; 

c) $m_n^B=\sum_T \pm m_{n, T}, n\ge 3$.

Here the summation is taken over all oriented planar trees $T$
with $n+1$ tails vertices (including the root vertex), such that 
the number of ingoing edges of every internal vertex of $T$
is at least $2$.
The linear map $m_{n, T}:B^{\otimes n}\to \Pi^nB$ where $\Pi$ stands for
parity reversion can be described in the following way. For every tree 
$T$
we consider an auxiliary tree $\bar{T}$ which is obtained
from $T$ by the insertion of a new vertex into
every internal edge. There will be two types of internal
vertices in ${\bar{T}}$: the ``old'' vertices, which coincide with the
internal vertices of $T$, and
the ``new'' ones, which can be thought geometrically as the midpoints of the
internal
edges of $T$. 

To every tail vertex of ${\bar{T}}$ we assign the embedding $i$.
To every ``old'' vertex $v$ we assign $m_k$ where $k$ stands for the
number of ingoing edges. To every ``new''
vertex we assign the homotopy operator $H$. To the root we assign the projector
$p$. Then moving along the tree down to the root one reads off the map $m_{n, T}$ as the composition of maps assigned
to vertices of $\bar{T}$.

An analog of $PQ$-manifold is an A$_{\infty }$-algebra with 
an odd innerproduct. Such an algebra can be considered as a linear space $E$ equipped
with such an odd inner product $<., .>$ that $<a_{0}, m_{k}(a_{1}, ..., a_{k})>=
\mu (a_{0}, a_{1}, ..., a_{k})$ is cyclically (graded)symmetric. If we have an $
A_{\infty }$-algebra specified by means of polylinear operators $
m_{k}(a_{1}, ..., a_{k})$ on $E$ then the graded symmetrization of $m_{k}$
specifies maps defining an L$_{\infty }$-algebra. An odd invariant inner
product on A$_{\infty }$-algebra can be considered as an invariant inner
product on the corresponding L$_{\infty }$-algebra. For any A$_{\infty }$
-algebra $E$ we can construct an A$_{\infty }$-algebra $E_{N}$
as a tensor product $E\otimes Mat_{N}$. More precisely a basis of $E_{N}$
consists of elements $(e_{k})_{\alpha }^{\beta }$ where $1\leq k\leq n$, $
1\leq \alpha , \beta \leq N$ and 

\begin{equation}\notag
m_{k}((e_{i_{k}})_{\alpha _{1}}^{\beta _{1}}, ..., (e_{i_{k}})_{\alpha
_{k}}^{\beta _{k}})=m_{i_{1}, ..., i_{1}}^{a}\delta _{\beta _{1}}^{\alpha
_{2}}\delta _{\beta _{2}}^{\alpha _{3}}...\delta _{\beta _{n-1}}^{\alpha
_{n}}\delta _{\beta }^{\alpha _{1}}\delta _{\beta _{n}}^{\alpha}(e_{a})_{\alpha }^{\beta }
\end{equation}

If the A$_{\infty }$-algebra algebra $E$ is equipped with an odd invariant
inner product the same is true for A$_{\infty }$-algebra $E_{N}$ and for
corresponding L$_{\infty }$-algebras $LE_{N}$.

\subsection*{Acknowledgments}
We would like to thank  A. Konechny, A. Losev, D. Piontkovski, L. 
Positselsky, A. Rosly, A. Waldron, E. Witten for 
stimulating discussions. Part of this work was done when one or both 
of the authors were staying in ESI, IHES, MPIM and 
Mittag-Leffler Institute; we  appreciates the 
hospitality of these institutions.

\end{document}